\documentclass[prb,preprint,superscriptaddress,showpacs,preprintnumbers,eqsecnum,amsmath,amssymb]{revtex4}
\usepackage{graphicx}
\usepackage{dcolumn}
\usepackage{bm}
\usepackage{color}
\usepackage{epsfig}

\begin{document}

\title{
Statistical study of the conductance and shot noise in 
open quantum-chaotic cavities: Contribution from whispering gallery modes}

\author{Evgeny N. Bulgakov}
\affiliation{Kirensky Institute of Physics, 660036, Krasnoyarsk, Russia}
\affiliation{Max-Planck-Institut f\"ur Physik komplexer Systeme,
D-01187 Dresden, Germany}

\author{V\'{\i}ctor A. Gopar}
\affiliation{Instituto de Biocomputaci\'on y F\'{\i}sica de Sistemas 
Complejos, Universidad de Zaragoza, Corona de Arag\'on, 42, 50009 
Zaragoza, Spain}
\affiliation{Max-Planck-Institut f\"ur Physik komplexer Systeme,
D-01187 Dresden, Germany}

\author{Pier A. Mello
\footnote{On leave from Instituto de F\'{\i}sica, U.N.A.M., Apartado Postal
20-364, 01000 M\'exico, D. F., M\'exico.}}
\affiliation{Max-Planck-Institut f\"ur Physik komplexer Systeme,
D-01187 Dresden, Germany}

\author{Ingrid Rotter}
\affiliation{Max-Planck-Institut f\"ur Physik komplexer Systeme,
D-01187 Dresden, Germany}


\begin{abstract}

In the past, a maximum-entropy model was introduced and applied to the study 
of statistical scattering by chaotic cavities, when short paths may play an important role in the scattering process.
In particular, the validity of the model
was investigated in relation with the statistical properties of the conductance 
in open chaotic cavities.
In this article we investigate further the validity of the maximum-entropy model,
by comparing the theoretical predictions with the results of computer simulations,
in which the Schr\"odinger equation is solved numerically inside the cavity
for one and two open channels in the leads;
we analyze, in addition to the conductance, the zero-frequency limit of the
shot-noise power spectrum.  
We also obtain theoretical results for the ensemble average of this last quantity,
for the orthogonal and unitary cases of the circular ensemble and an arbitrary number of channels.
Generally speaking, the agreement between theory and numerics is good. 
In some of the cavities that we study, short paths consist of whispering gallery 
modes, which were excluded in previous studies. 
These cavities turn out 
to be all the more interesting, as it is in relation with them that we 
found certain systematic discrepancies in the comparison with theory.
We give evidence that it is the lack of stationarity inside the energy 
interval that is analyzed, and hence the lack of ergodicity
--a property assumed in the maximum-entropy model-- 
that gives rise to the discrepancies. 
Indeed, the agreement between theory and numerical simulations is improved when the energy interval is reduced to a point and the statistics is then collected over an ensemble obtained by varying the position of an obstacle inside the cavity.
It thus appears that the maximum-entropy model is valid beyond the domain where it was originally derived. 
An understanding of this situation is still lacking at the present moment.

\end{abstract}

\maketitle

\section{Introduction}
\label{intro}

The statistical scattering of waves through open chaotic cavities
has been of great interest to many groups along the years 
\cite{dresden,mello-kumar}. The investigations that have been carried out 
are relevant to a variety of problems,
like the electronic transport through ballistic quantum dots, 
or the scattering of classical waves (e.g., electromagnetic 
or elastic waves) in chaotic billiards.

The approach provided by Random-Matrix Theory
has been particularly fruitful in the study of the statistical fluctuations of
transmission and reflection of waves by a number of systems, including 
billiards with a chaotic classical dynamics.
Within this approach we wish to focus our attention on 
the model of Refs. [\onlinecite{baranger-mello(epl),baranger-mello(wrm),mello-kumar}],
which was introduced originally in the context of Nuclear Physics
and was then applied to the domain of chaotic cavities.

We recall that, very generally, we can describe a scattering process in 
terms of a scattering matrix $S$. In the model referred to above, the 
statistical features of the problem are represented
by a measure in $S$-matrix space which, through the assumption 
of ``ergodicity", gives the probability of finding $S$ in a given volume 
element as the energy $E$ changes
and $S$ wanders through that space. The problem is, of course, to find 
that measure. The key assumption is made that in the scattering process two 
distinct time scales occur, associated, respectively, with 
a prompt, or direct, response due to the presence of short paths,
and a delayed, or equilibrated, response due to very long paths.
It turns out that the prompt, or direct, processes can be expressed in terms of the 
energy average of $S$, $\bar S$, also known as the {\em optical} $S$ matrix.
The statistical distribution of the scattering matrix $S$ is then constructed 
through a maximum-entropy ``ansatz",
assuming that it depends parametrically solely on the optical 
matrix. 
The notion of ergodicity, which allows replacing energy averages by 
ensemble averages, e.g., $\langle S \rangle= \bar S$, is essential to the argument.

The statistical properties of the conductance predicted by the maximum-entropy model we just described have been studied in the past;
these predictions have been also compared with the results of computer 
simulations which consist in solving the scalar Schr\"odinger equation 
numerically for a number of structures 
\cite{baranger-mello(epl),baranger-mello(wrm),mello-kumar}.
Although in those structures the two time scales referred to above were 
not as well separated as in Nuclear Physics problems, they seemed to us to 
be sufficiently distinct to allow a meaningful description.
It is the purpose of the present article 
to investigate further the validity of the maximum-entropy model,
by extending our earlier studies in the following three ways.

First, we wish to provide further predictions of our approach for other 
physical quantities in addition to the conductance.
For this purpose we analyze the zero-frequency limit of the shot-noise power spectrum $P$ at zero temperature.
For one open channel ($N=1$) we show that the problem can be reduced to quadratures and, in a number of cases, we can even study analytically
the influence of direct processes on the average, $\langle P \rangle$,
of the zero-frequency shot-noise power spectrum over an ensemble of cavities.
For an arbitrary number of channels, on the other hand, we show that
$\langle P \rangle$ 
can be evaluated analytically when direct processes are absent
[$\langle S \rangle =0$].

Second, we wish to extend the computer simulations mentioned above in a 
number of ways:

i) In some of the cavities used in the present paper the short paths 
consist of whispering gallery modes (WGM),
which were excluded in
Refs. [\onlinecite{baranger-mello(epl),baranger-mello(wrm)}] 
by the type of cavities that were used and the way the leads were attached. 
It is their effect  \cite{ingrid(2002),{wgm2}} that we wish to describe in terms 
of the optical $S$ matrix which, as we said, is precisely a measure of the 
short-time processes occurring in the scattering problem.
Information on the time scales involved could be provided by an analysis of 
the structure of $S(E)$ in the complex-energy plane.
Although we do not have direct access to the poles of the $S$ matrix, 
the complex eigenvalues of the so-called ``effective Hamiltonian" 
(which essentially consists of the Hamiltonian of the closed cavity plus 
the coupling to the continuum)
give evidence of a ``sea" of fine-structure, long-lived, resonances, 
plus a collection of shorter lived, more widely separated states.
This evidence is indicated in the present paper and studied in detail 
in Refs. [\onlinecite{ingrid_II, riddell, datta}].

ii) Earlier numerical simulations were performed for cavities with an applied magnetic field 
(the unitary universality class characterized by the Dyson 
parameter $\beta =2$),
in the presence of direct processes and for one channel ($N=1$).
The present simulations are performed for cavities with time-reversal invariance 
(the orthogonal universality class, characterized by the Dyson parameter 
$\beta =1$),
also in the presence of direct processes and for one ($N=1$) and two ($N=2$) open channels.

Third, we shall pay closer attention to the discrepancies between theory and numerical experiments.
Indeed, discrepancies similar to the ones that we shall observe in this paper were already
present, to a certain extent, in Ref. [\onlinecite{baranger-mello(epl)}], but were overlooked at that time.

The paper is organized as follows.
In the next section we first give a brief presentation of the
maximum-entropy model, recalling the assumptions that are used in its
derivation; 
these considerations will be important in the discussion to be presented in Sec. \ref{discussion}.
We then study a number of predictions of the model with regards to the
statistical properties of the conductance and the shot-noise power spectrum
at zero temperature.
In Sec. \ref{num} we present the results of the numerical simulations and 
the comparison with theory. 
Sec. \ref{num N1} is devoted to the one-channel case ($N=1$) 
and Sec. \ref{num N2} to two channels ($N=2$).
Finally, we discuss our results in Sec. \ref{discussion},
putting particular emphasis on  the discrepancies found between theory and numerical simulations.
We include an appendix, where some of the algebraic details of the relevant 
one- and two-channel statistical distributions are given.

\section{Statistical Model for the Description of Quantum Chaotic Scattering in Billiards}
\label{PK}

We present below the main ideas behind the maximum-entropy model briefly
described in the Introduction. This model was introduced in the past 
in the domain of Nuclear Physics and was later used to study the quantum
mechanical scattering occurring inside ballistic cavities 
(whose classical dynamics is chaotic)
connected to the outside by means of waveguides
\cite{baranger-mello(epl),baranger-mello(wrm),mello-kumar}.

The scattering problem can be described in terms of a scattering matrix $S$.
If the cavity is connected to two waveguides supporting $N$
channels each, the dimensionality of the $S$ matrix is $2N$.
As we mentioned in Sec. \ref{intro}, the model proposes a measure in 
$S$-matrix space which, through the assumption of ergodicity, describes 
the probability of finding $S$ in a given volume element as the energy 
$E$ changes and $S$ wanders through that space.
We write such a probability as
\begin{equation}
dP_{\left\langle S\right\rangle }^{(\beta )}(S)
=p_{\left\langle S\right\rangle }^{(\beta )}(S)d\mu _{\beta }(S) ,
\label{dP}
\end{equation}
where $p_{\left\langle S\right\rangle }^{(\beta )}(S)$,
referred to as the probability density, depends parametrically on the optical
matrix $\left\langle S\right\rangle$, as detailed below.
In the above equation, $d\mu ^{(\beta )}(S)$ is the {\em invariant measure} for the
universality class $\beta$
[we shall assume throughout that $\int d\mu _{\beta }(S)=1$].
Here we shall consider the cases $\beta=1$ (the orthogonal case) and 
$\beta=2$ (the unitary case), corresponding to cavities with and without time-reversal invariance, respectively, and in the absence of spin.
The problem is, of course, to find 
$p_{\left\langle S\right\rangle }^{(\beta )}(S)$.
To this end, a number of assumptions are made, as we now explain
(see Refs. [\onlinecite{mello-seligman,mello-pereyra-seligman}]).

1) The study of the statistical properties of $S(E)$ over an ensemble of 
cavities is simplified by idealizing $S(E)$, for real $E$, as a 
{\em stationary random (matrix) function} of $E$ satisfying the condition of {\em ergodicity}.

2) As explained in Sec. \ref{intro}, we assume that our scattering problem
can be characterized in terms of two time scales, arising from the prompt
and equilibrated components; 
the prompt response can be described in terms of the averaged $S$ matrix 
$\left\langle S\right\rangle$, also known as the {\em optical} $S$ matrix.

3) We assume $E$ to be far from thresholds, so that, locally, $S(E)$ 
is a meromorphic matrix function which is
analytic in the upper half of the complex-energy plane and has resonance 
poles in the lower half plane. 
From this follow what we have called in the past the ``analitycity-ergodicity" (AE) properties:
\begin{equation}
\big\langle \left( S_{a_{1}b_{1}}\right) ^{m_{1}}
\cdot \cdot \cdot
\left( S_{a_{k}b_{k}}\right) ^{m_{k}}\big\rangle
=\left\langle
S_{a_{1}b_{1}}\right\rangle ^{m_{1}}\cdot \cdot \cdot \left\langle
S_{a_{k}b_{k}}\right\rangle ^{m_{k}}.
\label{AE}
\end{equation}
This expression involves, on its left-hand side, only $S$ matrix
elements, whereas $S^{*}$ matrix elements are absent; on the right-hand
side, only the optical matrix $\left\langle S\right\rangle$
appears.
More generally, if $f(S)$ is a function that can be expanded as a series of non-negative powers of the
$S$ matrix elements, we must have the {\em reproducing property} \cite{hua}
\begin{equation}
\left\langle f(S) \right\rangle = f \left(  \left\langle S  \right\rangle \right) .
\label{reprod}
\end{equation}
One can then show that the probability density,
known as {\it Poisson's kernel},
\begin{equation}
p_{\left\langle S\right\rangle }(S)
=\frac
{[{\rm det}(I-\left\langle S\right\rangle \left\langle S\right\rangle ^{\dag})]
^{(2\beta N+2-\beta )/2}}
{|{\rm det}(I-S\left\langle S\right\rangle ^{\dag})|^{2\beta N+2-\beta }},
\label{poisson}
\end{equation}
is such that the average $S$ matrix is the optical matrix $\left\langle S\right\rangle$,
the AE requirements (\ref{AE}) and hence the reproducing property (\ref{reprod})
are satisfied, and the entropy
${\cal S}[p]$ associated with it,
$
{\cal S}[p]
\equiv -\int p_{\langle S\rangle }(S)\ln p_{\langle S\rangle}(S)d\mu (S) ,
\label{I[p]}
$
is greater than or equal to that of any other probability density
satisfying the AE requirements for the same $\left\langle S\right\rangle $.

With regards to the information-theoretic content of Poisson's kernel,
we have to distinguish between 
{\em i)} {\it general properties}, like 
unitarity of the $S$ matrix (flux conservation), 
analyticity of $S(E)$ implied by causality, 
and the presence or absence of time-reversal invariance (and spin-rotation
symmetry when spin is taken into account) which determines the
universality class (orthogonal, unitary or symplectic),
and 
{\em ii)} {\it particular properties}, parametrized by
the ensemble average
$\left\langle S\right\rangle$, 
which controls the presence of short-time processes.
System-specific {\it details other than the optical matrix $\langle S \rangle$ are assumed to be irrelevant.}
The optical matrix $\left\langle S\right\rangle$ is the only
``physically relevant parameter" assumed in the model.

From the probability distribution of Eqs. (\ref{dP}) and (\ref{poisson})
one can find the statistical properties  
of the quantities of interest over an ensemble of cavities.
In this paper we shall be concerned with the conductance and 
the zero-frequency shot noise power spectrum.

The dimensionless dc conductance [$g=G/(e^2/h)$] at zero temperature and for
the spinless case is 
given by Landauer's formula \cite{mello-kumar}
\begin{eqnarray}
\label{conductance}
g&=&T=tr(tt^{\dagger}) \nonumber\\
 &=&\sum_a \tau_a ,
\end{eqnarray}
where $\tau _a$ ($a=1,\cdots ,N$) are the eigenvalues of the Hermitean matrix 
$tt^{\dagger}$, and the transmission matrix $t$ is an $N\times N$ block of the 
$2N$-dimensional $S$ matrix which, in turn, is written as
\begin{equation}
S=\left[
\begin{array}{cc}
r &  t'   \\
t & r'
\end{array}
\right]\; .
\label{S}
\end{equation}

The zero-frequency limit of the shot-noise power spectrum at zero temperature can be 
expressed as
\cite{buettiker,carlo-rev}
\begin{equation}
P = P_0 \sum_{a=1}^N \tau _a (1-\tau _a),
\hspace{1cm} P_0=2eV\frac{2e^2}{h} .
\label{P}
\end{equation}
The average of $P$ over an ensemble of cavities will be written in the 
two alternative ways:
\begin{subequations}
\begin{eqnarray}
\langle P \rangle 
&=& P_0 \left\langle \sum_{n=1}^N \tau _a (1-\tau _a) \right\rangle
\label{<P> 1}
\\
&=& \langle P_P \rangle
\frac
{\left\langle\sum_{a=1}^N \tau _a (1-\tau _a)\right\rangle}{\langle \sum_{a=1}^N \tau _a\rangle} ,
\hspace{1cm}  \langle P_P \rangle=2eV\frac{2e^2}{h} \langle  T \rangle .
\label{<P> 2}
\end{eqnarray}
\label{<P>}
\end{subequations}
Here, $P_P$ is the result that would obtain if the noise were a Poissonian 
process, i.e., if there were no correlations among electrons and the electronic transport were completely random;
$T$ is the dimensionless conductance, Eq. (\ref{conductance}).
We see that since the shot-noise power is not determined simply by the conductance, it is only in the limit
$\tau _a \ll 1$ ($a=1,\cdots,N$) that we recover the Poissonian result.

It is clear that we need, for our purposes,
the joint probability distribution of the $\tau _a$'s. 
This can be found from Eq. (\ref{poisson}) as
\begin{subequations}
\begin{eqnarray}
w_{\langle S \rangle}^{(1)}(\tau _1, \dots, \tau _N)
&=&C_1 \frac{\prod _{a<b} \left| \tau _a - \tau _b  \right|}{\prod _c \sqrt{\tau _c}}
\int \int 
\frac{|\det (I-\langle S \rangle \langle S \rangle^{\dagger})|^{N+\frac12}}
{|\det (I-\langle S \rangle S^{\dagger})|^{2N+1}}
\;d\mu (v^{(1)})d\mu (v^{(2)})
\label{w(tau_a) beta1}
\nonumber \\
\\
w_{\langle S \rangle}^{(2)}(\tau _1, \dots, \tau _N)
&=&C_2 \prod _{a<b} \left( \tau _a - \tau _b  \right)^2
\int \cdots \int 
\frac{|\det (I-\langle S \rangle \langle S \rangle^{\dagger})|^{2N}}
{|\det (I-\langle S \rangle S^{\dagger})|^{4N}}
\;\prod _{i=1}^4 d\mu (v^{(i)}) \; ,
\nonumber \\
\label{w(tau_a) beta2}
\end{eqnarray}
\label{w(tau_a)}
\end{subequations}
for $\beta =1$ and $\beta =2$, respectively.
The quantity $C_{\beta }$ is a normalization constant.
The unitary matrices $v^{(i)}$ are the ones that occur in the polar decomposition of the $S$ matrix \cite{mello-kumar}
\begin{equation}
S =
\left[
\begin{array}{cc}
v^{(1)} &  0   \\
0 & v^{(2)}
\end{array}
\right] 
\left[
\begin{array}{cc}
-\sqrt{1-\tau} &  \sqrt{\tau}\\
 \sqrt{\tau}  & \sqrt{1-\tau}
\end{array}
\right]
\left[
\begin{array}{cc}
v^{(3)} &  0   \\
0 & v^{(4)}
\end{array}
\right] \; ,
\label{S polar}
\end{equation}
where $\tau$ stands for the $N\times N$ diagonal matrix constructed from the
the eigenvalues $\tau _a$ ($a=1,\cdots ,N$) of the Hermitian matrix 
$tt^{\dagger}$ [see Eq. (\ref{S})]
and the $v^{(i)}$ are arbitrary $N\times N$ unitary matrices for $\beta=2$,
with the restrictions $v^{(3)}=[v^{(1)}]^T$ and $v^{(4)}=[v^{(2)}]^T$
for $\beta=1$.

In what follows we study, in particular, the cases in which the two waveguides 
connecting the cavity 
to the outside may support one, two, or an arbitrary number of open channels.

\subsection{The one-channel case, $N=1$}
\label{N=1}

In this case we have only one $\tau $, which coincides with the 
conductance $T$, whose probability distribution can thus be written 
from Eqs. (\ref{w(tau_a)}) as 
\begin{subequations}
\begin{eqnarray}
w_{\langle S \rangle}^{(1)}(T)
&=&\frac{1}{2\sqrt{T}}
\int _0 ^{2\pi} \int _0 ^{2\pi}
\frac{|\det (I-\langle S \rangle\;\langle S \rangle^{\dagger})|^{3/2}}
{|\det (I-\langle S \rangle S^{\dagger})|^3}
\;\frac{d\alpha d\beta}{(2\pi)^2} ,
\label{w(g) beta1 N=1}
\\
w_{\langle S \rangle}^{(2)}(T)
&=& \int _0 ^{2\pi}\cdots \int _0 ^{2\pi}
\frac{|\det (I-\langle S \rangle\;\langle S \rangle^{\dagger})|^{2}}
{|\det (I-\langle S \rangle S^{\dagger})|^4}
\;\frac{d\alpha d\beta d\gamma d\delta}{(2\pi)^4} .
\label{w(g) beta2 N=1}
\end{eqnarray}
\label{w(g) beta12 N=1}
\end{subequations}
The polar representation of $S$ for $N=1$ is written down explicitly in 
Eq. (\ref{S polar N=1}) of the Appendix.

In the absence of direct processes, i.e., $\langle S \rangle =0$, the 
$T$ distribution
of Eqs. (\ref{w(g) beta12 N=1}) reduces to the well known results
\begin{subequations}
\begin{eqnarray}
w_0^{(1)}(T)&=&\frac{1}{2\sqrt{T}},
\label{w(g) beta1 N=1 S0}
\\
w_0^{({2})}(T)&=&1 ,
\label{w(g) beta2 N=1 S0}
\end{eqnarray}
\label{w(g) N=1 S0}
\end{subequations}
for the orthogonal ($\beta =1$) and unitary ($\beta =2$) cases, 
respectively.

The $T$ distribution for the unitary case, Eq. (\ref{w(g) beta2 N=1}), can 
be integrated explicitly \cite{mello-kumar}. 
As an example, for the particular case 
$\langle t \rangle=\langle t' \rangle=0$, corresponding to 
direct reflection and no direct transmission,
and assuming, for simplicity, the ``equivalent-channel" case
($|\langle r \rangle| =|\langle r' \rangle|$), one finds
\begin{eqnarray}
w_{\langle r \rangle}^{(2)}(T)
=(1-| \langle r  \rangle |^2)
\frac
{(1-| \langle r  \rangle |^4)^2 + 2| \langle r  \rangle |^2(1+| \langle r  \rangle |^4)T + 4| \langle r  \rangle |^4T^2}
{\left[(1-| \langle r  \rangle |^2)^2 + 4 | \langle r  \rangle |^2 T\right]^{5/2}}  .
\label{w(g) beta2 N=1 anal}
\end{eqnarray}
For the case of direct transmission and no direct reflection, the result is 
obtained from the previous equation by replacing 
$| \langle r  \rangle |$ by $|\langle t \rangle|$ and $T$ by $1-T$.

\begin{figure}[t]
\begin{center}
\includegraphics[width=0.6\columnwidth]{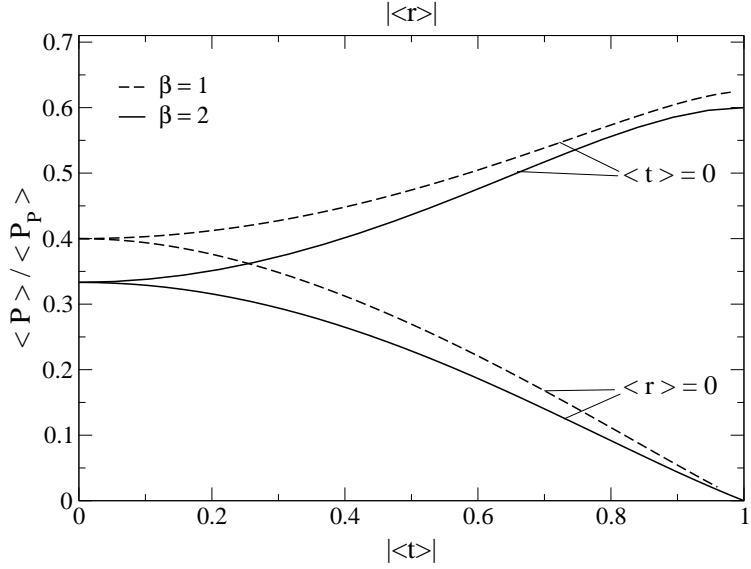}
\end{center}
\caption
{The ratio
$\langle P \rangle / \langle P_P \rangle$ 
as a function direct reflection $|\langle r \rangle|=|\langle r' \rangle|$
(indicated in the upper horizontal line as the abscissa)
for the case $\langle t \rangle =\langle t' \rangle =0$
is shown as the two upper curves.
The two lower curves show the same ratio as a function direct transmission
$|\langle t \rangle|=|\langle t' \rangle|$
(indicated in the lower horizontal line as the abscissa)  
for the case $\langle r \rangle =\langle r' \rangle =0$.
The dashed lines correspond to the orthogonal universality class ($\beta =1$)
and the solid lines to the unitary class ($\beta =2$).
}
\label{fanobeta2N1}
\end{figure}

The $\tau ( =T)$ distribution for the unitary case given in 
Eq. (\ref{w(g) beta2 N=1 anal})
allows us to study the effect of direct processes on the averaged shot-noise 
power spectrum $\langle P \rangle$ of Eq. (\ref{<P> 2}); 
this case is particularly suited to gain some physical insight, since
the result for $\langle P \rangle$ can be expressed analytically in a 
remarkably simple fashion. 
For the particular case of direct reflection  and no
direct transmission ($\langle t \rangle= \langle t' \rangle=0$), and assuming 
$|\langle r \rangle| = |\langle r' \rangle|$, one finds, 
from Eq. (\ref{w(g) beta2 N=1 anal}), the result:
\begin{equation}
\label{direct_r}
\frac{\langle P \rangle^{(2)}}{\langle P_P \rangle^{(2)}}
=\frac{1}{5} \;
\frac{  5 - 9|\langle r \rangle |^4 + 4|\langle r \rangle |^6 }
{ 3 -4|\langle r \rangle|^2 + |\langle r \rangle|^4},
\end{equation}
while for direct transmission and no direct reflection
($\langle r \rangle= \langle r' \rangle=0$),
and assuming $|\langle t \rangle| = |\langle t' \rangle|$, one obtains
\begin{equation}
\label{direct_t}
\frac{\langle P \rangle^{(2)}}{\langle P_P \rangle^{(2)}}
=\frac{1}{5} \;
\frac{  5  -9|\langle t \rangle |^4 + 4|\langle t \rangle |^6}
{  3 +4|\langle t \rangle|^2 - |\langle t \rangle|^4  }.
\end{equation}
In Fig. \ref{fanobeta2N1} the behavior of the ratio
$\langle P \rangle^{(2)}/\langle P_P \rangle^{(2)}$ as a
function of $|\langle r \rangle|=|\langle r '\rangle|$ for the direct reflection case
($\langle t \rangle =\langle t ' \rangle =0$), 
Eq. (\ref{direct_r}), is shown as the upper solid curve;
the lower solid curve shows the case of direct transmission as a
function of $|\langle t \rangle |=|\langle t' \rangle |$ 
(when $\langle r \rangle =\langle r' \rangle =0$), 
Eq. (\ref{direct_t}). 
For the upper curve, 
the ratio $\langle P \rangle ^{(2)}/ \langle P_P \rangle^{(2)}$
increases as a function of $|\langle r \rangle |$;
since, as $|\langle r \rangle | \to 1$,
$w(T) \to \delta (T)$,
at first sight one would expect, in this limit, the ratio 
$\langle P \rangle / \langle P_P \rangle^{(2)}$ to increase 
towards the Poissonian value unity.
That this is not the case is due to the fact that both 
$\langle T \rangle$ and $\langle T^2 \rangle$
tend to zero linearly with $1- |\langle r \rangle |$ as this quantity 
tends to zero.

For the orthogonal symmetry class ($\beta=1$) we have not succeeded in finding an analytical expression for the conductance distribution, even for the particular cases studied above.
For these cases, the ratio 
$\langle P \rangle^{(1)}/\langle P_P \rangle^{(1)}$ 
was thus calculated numerically from Eq. (\ref{w(g) beta1 N=1})
and the results are also presented in
Fig. \ref{fanobeta2N1} for comparison with the unitary case; 
we observe that the ratio
$\langle P \rangle^{(\beta )}/\langle P_P \rangle^{(\beta )}$ 
is always larger for $\beta=1$ than for $\beta=2$.

We wish to point out a property of the average shot-noise power 
$\langle P \rangle^{(\beta )}$ of Eq. (\ref{<P> 1}), in the present one-channel case.
Poisson's kernel of Eq. (\ref{poisson}) has the property that has been called ``covariance'' \cite{mello-pereyra-seligman}:
$p_{\langle S \rangle}^{(\beta )}(S)
= p_{\langle \tilde S \rangle}^{(\beta )}(\tilde S)$,
where 
$\tilde S =U_0 S V_0$,
$U_0$ and $V_0$ being fixed unitary matrices for $\beta=2$, with 
$V_0=U_0^T$ for  $\beta=1$, the same transformation being applied to the optical
$\langle S \rangle$.
The invariant measure is invariant under this transformation.
For $\beta =2$, one can verify that the unitary 
matrices 
$U_0=\left[
\begin{array}{cc}
0 &  1  \\
1 & 0 \end{array}
\right]$ and 
$V_0={\mathbf 1}$ 
exchange $r \ (r')$ and $t \ (t')$ and 
their corresponding average values appearing in $\langle S \rangle$.
For the case $\beta=1$, we have $t=t'$. 
If we also have $r=r'$, as in the case of a system with ``left-right symmetry'',
the matrix 
$U_0=\frac{1}{\sqrt{2}}\left[
\begin{array}{cc}
1+i &  1-i  \\
1-i & 1+i \end{array} \right]$ 
switches $r$ and $t$ 
and the corresponding optical parameters.
The above transformations keeps $P=P_0 \tau (1-\tau )$ invariant. As a consequence, $\langle P \rangle^{(\beta )}$ remains invariant under the interchange
$\langle r \rangle \leftrightarrow \langle t \rangle$,
$\langle r' \rangle \leftrightarrow \langle t' \rangle$
for $\beta =2$, and $\langle r \rangle \leftrightarrow\langle t \rangle$ for the particular $\beta =1$ case mentioned above. 
We observe that, indeed, the numerators of Eqs. (\ref{direct_r}) and
(\ref{direct_t}), which are proportional to $\langle P \rangle^{(2 )}$, 
do fulfill this property. However, for the $\beta =1$ case considered here, this 
symmetry does not apply.

In the present one-channel case one can write down an expression for 
the distribution of the ``dimensionless" shot-noise power spectrum 
[see Eq. (\ref{P})]
\begin{equation}
\eta\equiv \frac{P}{P_0}= \tau(1-\tau) \; ,
\label{eta}
\end{equation}
which lies in the range $0 \leq \eta  \leq \frac14$
(we are using the notation of Ref. \onlinecite{buettiker}).
Since $\eta$ is a function of the conductance, we can make an elementary change of variables and
write
\begin{subequations}
\begin{eqnarray}
w^{(\beta )}(\eta)=\left[\frac{w^{(\beta )}(\tau)}{|1-2\tau|}\right]_{\tau = \tau(\eta)}
\\
\tau=\frac12\left[ 1\pm \sqrt{1-4\eta} \right] .
\end{eqnarray}
\label{w(eta)}
\end{subequations}
Thus the distribution in question is given by:
\begin{equation}
w^{(\beta)}(\eta) = \frac
{w^{(\beta)}(\tau=\frac{ 1 + \sqrt{1-4\eta}}{2}) + w^{(\beta)}(\tau=\frac{ 1 - \sqrt{1-4\eta}}{2})}
{\sqrt{1-4\eta}} ,
\label{w(eta)}
\end{equation}
where $w^{(\beta)}(\tau)$ is given in Eqs. (\ref{w(g) beta12 N=1}).
For $\langle S \rangle =0$, the result of this last equation (\ref{w(eta)})
reduces to Eq. (95) of Ref. \onlinecite{buettiker}.

\subsection{The two-channel case, $N=2$}
\label{N=2}

In the two-channel case the matrix $tt^{\dagger}$ is two-dimensional and has two
eigenvalues $\tau _1, \tau _2$, whose joint probability distribution can 
be written from Eqs. (\ref{w(tau_a)}) as

\begin{subequations}
\begin{eqnarray}
w_{\langle S \rangle}^{(1)}(\tau _1, \tau _2)
&=& \frac 34 \frac{|\tau _1 - \tau _2|}{\sqrt{\tau _1 \tau _2}}
\int \int
\frac
{[{\rm det}(I-\left\langle S\right\rangle \left\langle S\right\rangle ^{\dag})]^{5/2}}
{|{\rm det}(I-S\left\langle S\right\rangle ^{\dag})|^{5}}
d\mu (v^{(1)})d\mu (v^{(2)}) 
\label{w(tau) beta1 N=2}
\nonumber \\
\\
w_{\langle S \rangle}^{(2)}(\tau _1, \tau _2)
&=& 6 (\tau _1 - \tau _2)^2
\int \cdots \int
\frac
{[{\rm det}(I-\left\langle S\right\rangle \left\langle S\right\rangle ^{\dag})]^{4}}
{|{\rm det}(I-S\left\langle S\right\rangle ^{\dag})|^{8}}
d\mu (v^{(1)}) \cdots d\mu (v^{(4)}) .
\label{w(tau) beta2 N=2}
\nonumber \\
\end{eqnarray}
\label{w(tau) beta12 N=2}
\end{subequations}
Here, $d\mu (v^{(i)})$ is the invariant measure for the unitary matrices
$v^{(i)}$ used to represent $S$ in its polar form, Eq. (\ref{S 2});
the explicit form of $d\mu (v^{(i)})$ is given in
Eqs. (\ref{dmu vi}) and (\ref{range}).

From the above expressions we can evaluate the probability distribution of
the conductance as
\begin{equation}
w_{\langle S \rangle}^{(\beta)}(T)= \int _0 ^1
w_{\langle S \rangle}^{(\beta)}(\tau _1, T - \tau _1) d\tau _1 ,
\label{wT N=2}
\end{equation}
and the ratio $\langle P  \rangle  /  \langle P_P\rangle $
for the shot-noise power spectrum as
\begin{equation}
\frac{\langle P  \rangle}{\langle P_P \rangle}
=\frac{\langle \sum _{a=1}^2 \tau_a(1-\tau_a )\rangle}{\langle \sum _{a=1}^2 \tau_a\rangle} .
\label{P/PP N2}
\end{equation}
In the absence of direct processes, $\langle S \rangle =0$,
we obtain for $w_{0}^{(\beta )}(\tau _1, \tau _2)$ the well known results \cite{mello-kumar}:
\begin{subequations}
\begin{eqnarray}
w_{0}^{(1)}(\tau _1, \tau _2)
&=& \frac 34 \frac{|\tau _1 - \tau _2|}{\sqrt{\tau _1 \tau _2}}
\\
w_{0}^{(2)}(\tau _1, \tau _2)
&=& 6 (\tau _1 - \tau _2)^2 ,
\end{eqnarray}
\label{w(tau) beta12 N=2 S=0}
\end{subequations}
and for the conductance distribution $w_{0}^{(\beta )}(T)$
\begin{subequations}
\begin{eqnarray}
w_{0}^{(1)}(T) 
&=& 
\left\{    
\begin{array}{cc}
\frac 32 T, & \;\;\; 0<T<1
\\
\frac 32 \left(T-2\sqrt{T-1}\right), & \;\;\; 1<T<2 .
\end{array}
\right.
\\
w_{0}^{(2)}(T) 
&=& 
2\left[ 1 - \left|1 - T  \right|  \right]^3  .
\end{eqnarray}
\label{w(T) beta12 N=2 S=0}
\end{subequations}

\subsection{The case of arbitrary $N$}
\label{N arb}

In the absence of direct processes, $\langle S \rangle = 0$, 
various results concerning the average and variance of the conductance
are known \cite{mello-kumar} and will not be reproduced here.

Not known, to our knowledge, is the behavior of the shot-noise power spectrum for arbitrary $N$,
even for $\langle S \rangle = 0$.
We calculate below, for such a situation, the average 
$\langle P \rangle$ for the orthogonal and the unitary cases.

The numerator of (\ref{<P> 2}) can be written as
\begin{eqnarray}
\left\langle \sum_{a=1}^N \tau _a (1-\tau _a) \right\rangle_0^{(\beta)}
&=& \left\langle tr(tt^{\dagger})\right\rangle
- \left\langle tr(tt^{\dagger}tt^{\dagger})\right\rangle_0^{(\beta)}
\nonumber\\
&=&\sum _{a,b =1}^{N}\langle     |t_{ab}|^2   \rangle _0^{(\beta)}
- \sum _{a,b,c,d =1}^{N}\langle  t_{ab} t_{cd}t_{cb}^{\ast} t_{ad} ^{\ast}\rangle_0^{(\beta)}
\nonumber \\
&=&\sum _{a,b =1}^{N}\langle  S_{ab}^{21}\left[S_{ab}^{21}\right]^{\ast}  \rangle _0^{(\beta)}
-\sum _{a,b,c,d =1}^{N}\left\langle  S_{ab}^{21} S_{cd}^{21}\left[S_{cb}^{21} S_{ad}^{21}\right]^{\ast}\right\rangle_0^{(\beta)}
\label{F0}
\end{eqnarray}
The notation $\langle \cdots \rangle _0^{(\beta)}$ indicates an average over the invariant measure for the universality class $\beta$.

In the last line of Eq. (\ref{F0}) the upper indices $21$ indicate the $21$ block of the $S$ matrix in Eq. (\ref{S}).

Averages of monomials of the type
\begin{equation}
Q_{\alpha_1 \beta_1 , \cdots, \alpha_p \beta_p}^{\alpha'_1 \beta'_1 , \cdots, \alpha'_p  \beta'_p}
(\beta)
\equiv\left\langle S_{\alpha_1 \beta_1} \cdots  S_{\alpha_p \beta_p}
\left[S_{\alpha'_1 \beta'_1} \cdots  S_{\alpha'_p \beta'_p} \right]^{\ast}  \right\rangle_0^{(\beta)}
\label{M}
\end{equation}
were studied in Ref. \onlinecite{mello-seligman} and \onlinecite{mello-jpa}, for $\beta =1$
and $\beta =2$, respectively.
We now consider these two cases separately.

In the orthogonal case, $\beta =1$,
we denote $Q(1)\equiv M$, just as in Ref. [\onlinecite{mello-seligman}].
In that reference one finds the results
\begin{subequations}
\begin{eqnarray}
M_{\alpha \beta}^{\alpha' \beta'}
&=&\frac{\Delta_{\alpha \beta}^{\alpha' \beta'} + \Delta_{\alpha \beta}^{ \beta'\alpha'}}{2N+1}\; ,
\hspace{1cm}\Delta_{\alpha \beta}^{\alpha' \beta'}=\delta_{\alpha}^{\alpha '}\delta_{\beta} ^{\beta '}.
\label{M 2}
\\
M_{\alpha \beta,\gamma \delta}^{\alpha' \beta', \gamma' \delta '}
&=&A\left[M_{\alpha \beta}^{\alpha' \beta'}M_{\gamma \delta}^{\gamma' \delta'}
+ M_{\alpha \beta}^{\gamma' \delta'} M_{\gamma \delta}^{\alpha' \beta'} \right]
\nonumber \\
&&+B\left[M_{\alpha \beta}^{\alpha' \gamma'}M_{\gamma \delta}^{\beta' \delta'}
+ M_{\alpha \beta}^{\beta' \delta'} M_{\gamma \delta}^{\alpha' \gamma'}
+M_{\alpha \beta}^{\alpha' \delta'}M_{\gamma \delta}^{\beta ' \gamma' }
+ M_{\alpha \beta}^{\beta '\gamma' } M_{\gamma \delta}^{\alpha' \delta'}   \right],
\label{M 4}
\end{eqnarray}
where
\begin{equation}
A=\frac{(2N+1)(2N+2)}{2N(2N+3)},\hspace{1cm} B=-\frac{2N+1}{2N(2N+3)} .
\label{AB}
\end{equation}
\label{M 24}
\end{subequations}
Substituting the results (\ref{M 24}) in Eq. (\ref{F0})
we find for the average of $P$, Eq. (\ref{<P> 2}), for the 
orthogonal case:
\begin{eqnarray}
\langle P \rangle ^{(\beta =1)}
&=&\frac{(N+1)^2}{2N(2N+3)}
\;\langle P_P \rangle ^{(\beta =1)} \; .
\label{<P> beta1}
\end{eqnarray}

In the unitary case, $\beta =2$, we denote $Q(2)\equiv Q$, just as in
Ref. [\onlinecite{mello-jpa}].
In that reference one finds the results
\begin{subequations}
\begin{eqnarray}
Q_{\alpha \beta}^{\alpha' \beta'}
&=&\frac{\Delta_{\alpha \beta}^{\alpha' \beta'}}{N}\; ,
\label{Q 2}
\\
Q_{\alpha \beta,\gamma \delta}^{\alpha' \beta', \gamma' \delta '}
&=&\frac{1}{(2N)^2 -1}\left[
\Delta_{\alpha \gamma }^{\alpha' \gamma' }\Delta_{\beta \delta}^{\beta' \delta'}
+ \Delta_{\alpha \gamma }^{\gamma' \alpha'}\Delta_{\beta \delta}^{\delta' \beta'}
\right]
\nonumber \\
&&-\frac{1}{2N[(2N)^2 -1]}\left[
\Delta_{\alpha \gamma }^{\alpha' \gamma' }\Delta_{\beta \delta}^{\delta' \beta' }
+ \Delta_{\alpha \gamma }^{\gamma' \alpha'}\Delta_{\beta \delta}^{\beta' \delta' }
\right]
\label{M 4}
\end{eqnarray}
\end{subequations}
which has to be substituted in Eq. (\ref{F0}).
For $\langle P \rangle ^{(\beta =2)}$, Eq. (\ref{<P> 2}), we find:
\begin{equation}
\langle P \rangle ^{(\beta =2)}
= \frac{N^2}{4N^2 - 1} \; \langle P_P \rangle ^{(\beta =2)} .
\label{<P> beta2}
\end{equation}

For a large number of open channels, $N\gg 1$, Eqs. (\ref{<P> beta1}) and 
(\ref{<P> beta2}) give
$\langle P \rangle ^{(\beta )} 
\approx \frac 14 \langle P_P \rangle ^{(\beta )}
\approx  NP_0/8 $, 
just as in Refs. [\onlinecite{buettiker,carlo-rev}].

\begin{figure}[h]

\epsfig{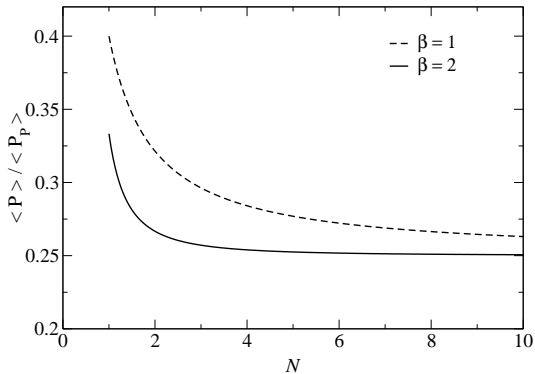}

\caption{The ratio $\langle P \rangle / \langle P_P \rangle$ for $\beta =1$ (upper curve)
and $\beta =2$ (lower curve) for $\langle S \rangle = 0$, as a function of the number of channels $N$.
}
\label{fanobeta12 N}
\end{figure}

The ratio $\langle P \rangle ^{(\beta)}/\langle P_P \rangle ^{(\beta)}$
from Eqs. (\ref{<P> beta1}) and (\ref{<P> beta2}) is plotted in Fig. 
\ref{fanobeta12 N} as a function of the number of channels $N$.
We observe that this ratio is always larger for the orthogonal 
($\beta =1$) than for the unitary case ($\beta =2$),
just as was noticed in the results shown in Fig. \ref{fanobeta2N1}
for the one-channel case.
This effect indicates that time reversal symmetry pushes 
the $\tau _a$ distribution towards small $\tau _a$'s
[for $N=1$ this effect is given by Eq. (\ref{w(g) beta1 N=1 S0})]
in such a way that
$\langle P \rangle ^{(\beta =1)}$ gets closer to Poisson's value.

\section{Numerical Simulations}
\label{num}

The maximum-entropy approach that we have been discussing is expected to be valid for cavities in which the classical dynamics is completely chaotic 
--a property that refers to the long-time behavior of the system--
as in such structures the long-time response is equilibrated and classically ergodic.

In Refs. [\onlinecite{baranger-mello(epl),baranger-mello(wrm)}] the 
scalar Schr\"odinger equation was integrated numerically for a number of 2D cavities in order to examine to what extent our approach really holds.
In those references the analysis was performed for the conductance distribution $w(T)$.
The cavities were subjected to a magnetic field ($\beta =2$) and they were connected to the outside by waveguides admitting one open channel ($N=1$). 
Moreover, 
the structures were such that they obviously supported short paths associated with direct reflection from a barrier, direct transmission from one lead to the other, or skipping-orbit trajectories in the presence of the magnetic field.

In what follows we consider the numerical solution of the Schr\"odinger equation for
2D structures which again support direct processes.
Now the system is not immersed in a magnetic field, so that it is time-reversal invariant
($\beta =1$).
We mainly study the one-channel case, $N=1$
(Sec. \ref{num N1} below),
although we also present some results for $N=2$ (Sec. \ref{num N2}).

In addition to the conductance distribution, the average of the zero-frequency shot noise 
power spectrum is also studied, in order to examine further the applicability of the model.
Ensembles of similar systems are obtained by introducing an obstacle inside the cavity and 
changing its position (see Figs. \ref{w(T)22-23}, \ref{wgm} and \ref{w(T) 75} below).
In all cases the optical $S$ matrix $\langle S \rangle$ was extracted from the data
and used as an input in Eq. (\ref{poisson}), or in the various results of the
Sec. \ref{PK}, to produce the theoretical predictions to be compared with
numerical experiments. 
In this sense all of our fits are ``parameter free". For details of the
numerical study see Refs.  [\onlinecite{ingrid_II,riddell,datta}].

\subsection{The one-channel case, $N=1$}
\label{num N1}

When the energy $E$ lies inside the interval
$\frac{\hbar^2}{2mW^2}\left[ N^2\pi ^2, \;\; (N+1)^2 \pi ^2 \right]$,
each waveguide (of width $W$) supports exactly $N$ open channels.
In units such that $\hbar^2/2mW^2 =1$, one open channel ($N=1$) occurs for
$E \; \epsilon \; [\pi^2, 4\pi^2]\approx [10,40]$.
We need to study $S(E)$ in energy intervals $\Delta E$ not too close to either threshold, in order to avoid threshold singularities.

\subsubsection{Statistical properties of the conductance}
\label{num cond N=1}

\begin{widetext}
\begin{figure}[h]
\epsfig{file=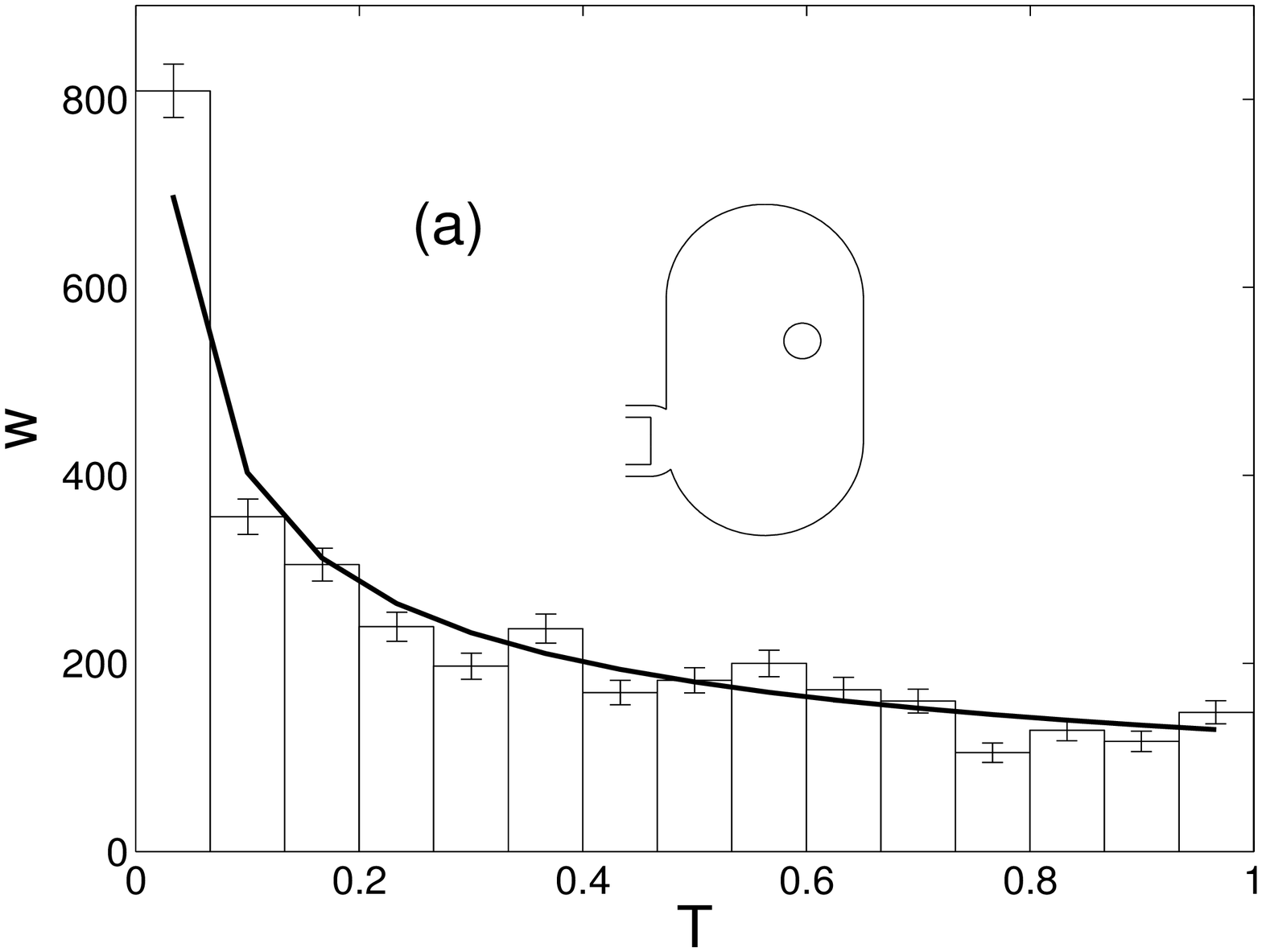,width=6cm,clip=}
\epsfig{file=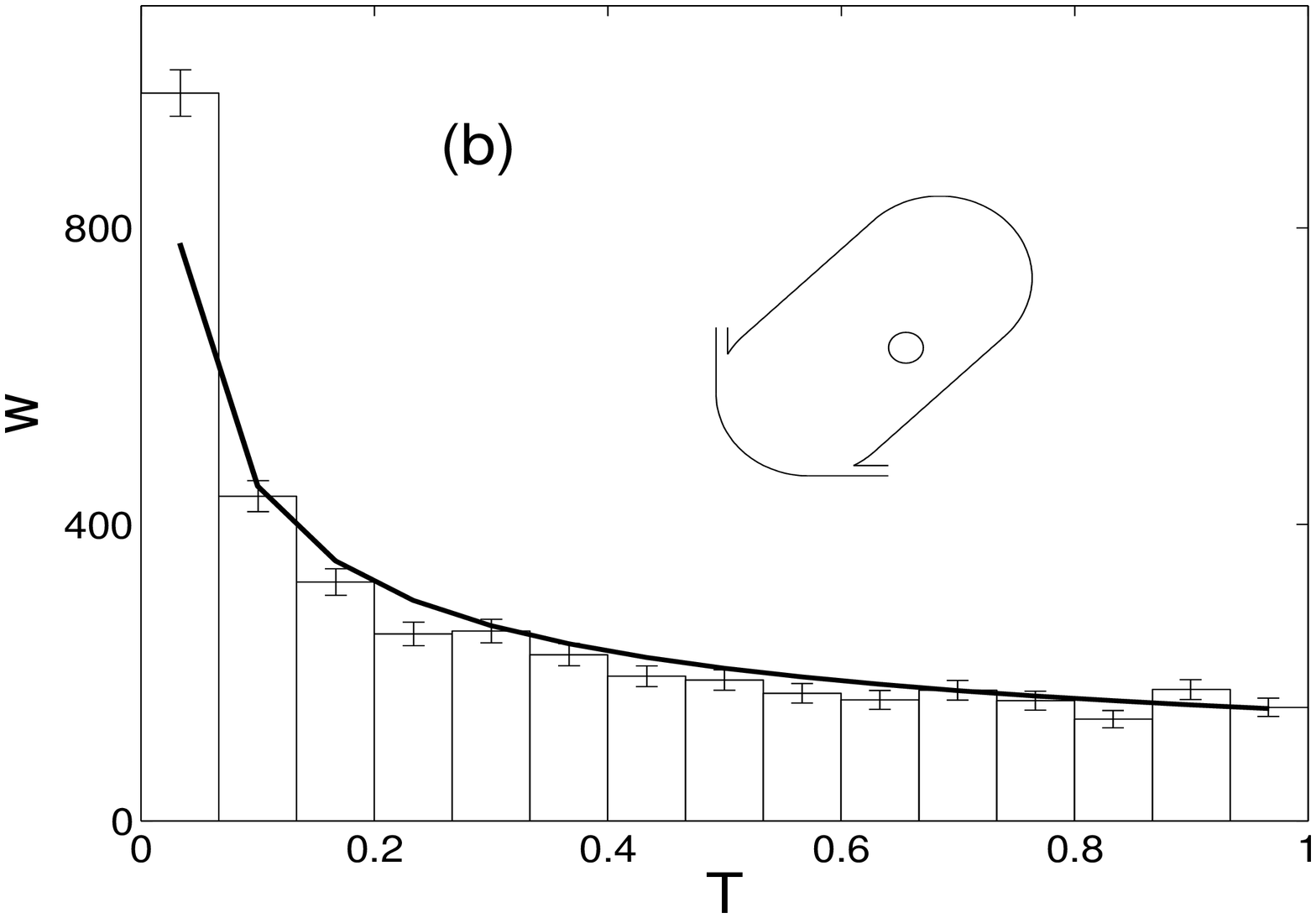,width=6cm,clip=}
\epsfig{file=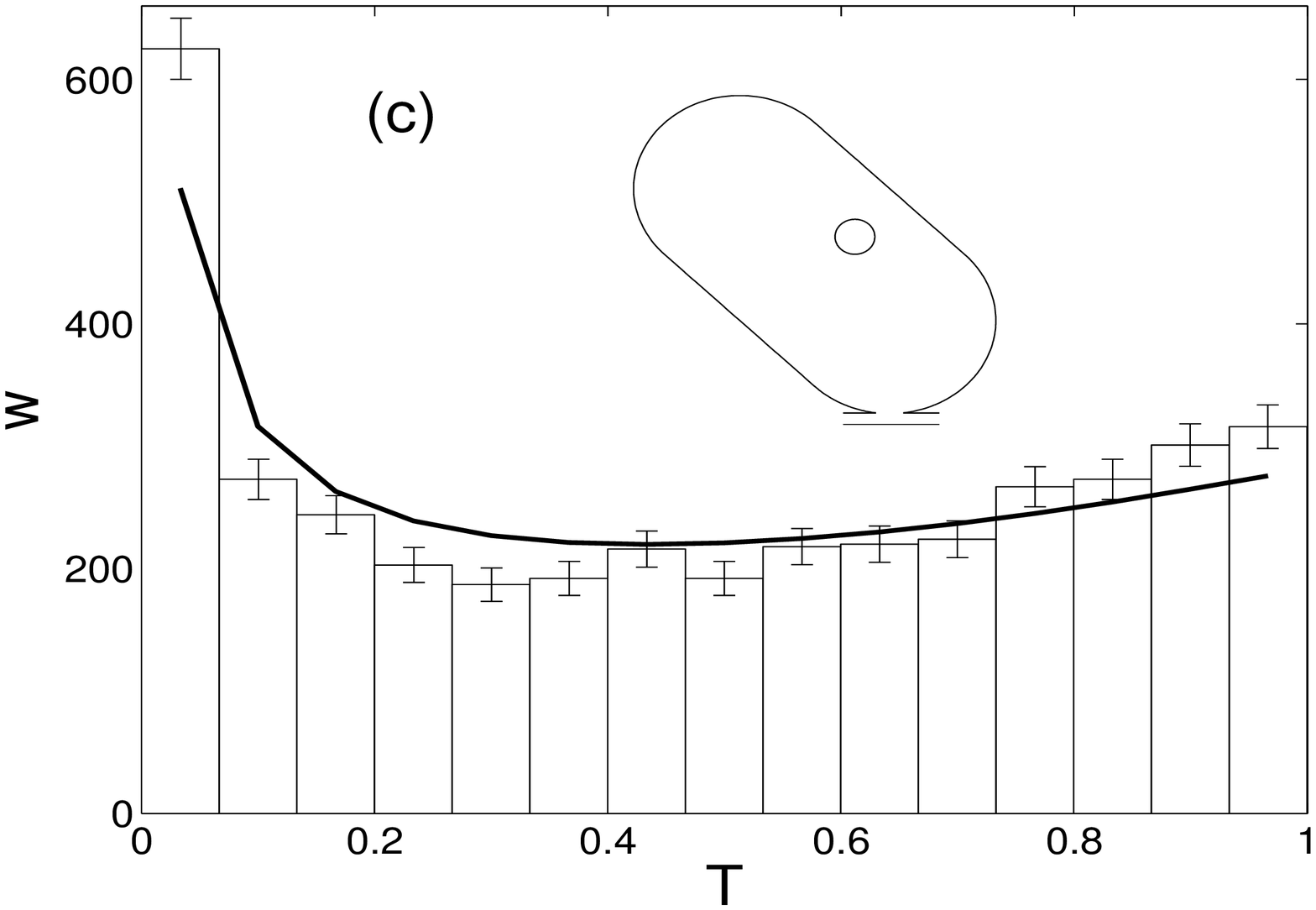,width=6cm,clip=}
\epsfig{file=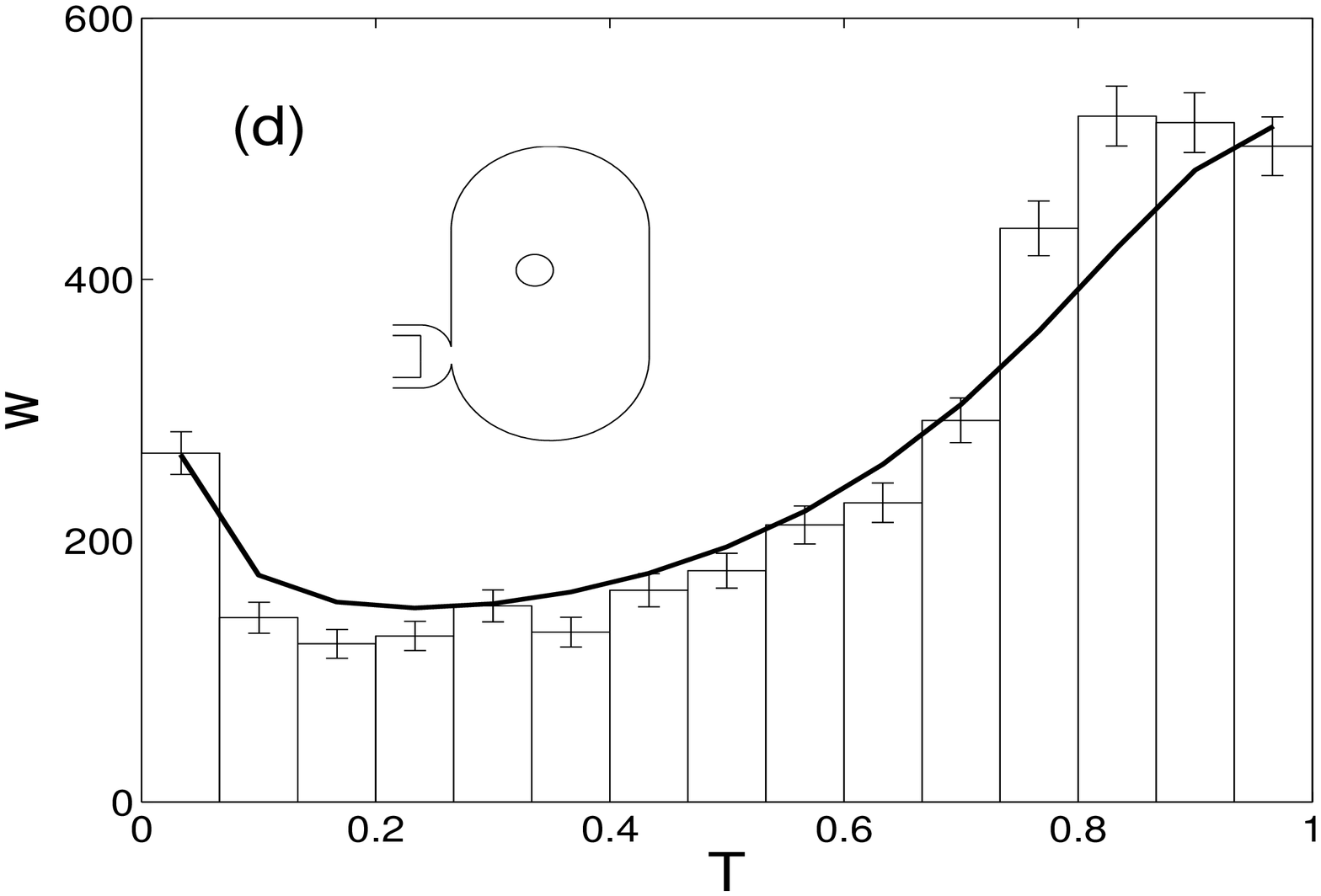,width=6cm,clip=}
\caption{{\footnotesize 
The distribution $W$ (not be confused with the width of waveguide $W$ used
in the text) of the conductance, normalized to the total number of cases,
for the structures shown in the insets.
Each bin shows the frequency that occurred in that interval.
The histograms were obtained from a numerical solution of the Schr\"odinger equation in the
energy interval $E \in [22,23]$, as explained in the text.
The theoretical distributions, obtained from Poisson's kernel using the
optical $S$ matrix extracted in each of the four cases,
were calculated at 15 points (in the interval $0<T<1$) which were then joined to obtain 
the continuous curves.
The agreement between theory and numerical simulations is, in general, good;
the largest deviations occur in panel (d), where the optical, or direct, transmission, 
is the largest, due to whispering gallery modes in the small cavity.
}}
\label{w(T)22-23}
\end{figure}
\end{widetext}

Fig. \ref{w(T)22-23} shows, as insets, the structures for which the numerical study was performed:
they consist of a Bunimovich stadium connected to two waveguides directly, as in panels (a), (b) and (c), or through a smaller half stadium, as in (d).
The structures are spatially asymmetric.

The histograms were obtained by solving the Schr\"odinger equation inside these structures and 
collecting the data in the energy interval
$\Delta E =[22,23]$ (in the units explained above), and then across an ensemble of 200 positions
of the obstacle, which is also shown in the figure.
In that energy interval, 20 equally-spaced points were considered: 
these points are farther apart than the correlation energy, as it appears from the negligible 
correlation coefficient (over the ensemble) that was obtained for the transmission and 
reflection amplitudes for two successive points.
The optical $S$ matrix, obtained as an energy plus an ensemble average of $S$, i.e., 
$\langle \bar{S}\rangle$,
was extracted from the data and the optical reflection and transmission matrix elements are 
given in Table \ref{S-opt-fig3}.


\begin{table}[h]
\caption{\footnotesize The optical reflection and transmission matrix elements
for the four cases in Fig. \ref{w(T)22-23}.}
\begin{tabular}{||cccc||}
\hline\hline
{Case} &  \multicolumn{1}{|c}
{$
\langle \bar{r}\rangle
$} &  \multicolumn{1}{|c}
{$\langle \bar{t}\rangle = \langle \bar{t'}\rangle $
}  
&\multicolumn{1}{|c||} 
{$
\langle \bar{r'}\rangle
$}  \\
\hline\hline
{Fig. 3a} &\multicolumn{1}{|c}{0.0007-0.0651i}&  \multicolumn{1}{|c}{-0.0725+0.0078i} &\multicolumn{1}{|c||}{0.0040+0.0008i}  \\
\hline
{Fig. 3b} &\multicolumn{1}{|c}{-0.0384-0.0213i}&  \multicolumn{1}{|c}{-0.0767-0.0211i}&\multicolumn{1}{|c||}{-0.0388-0.0210i}  \\ 
\hline
{Fig. 3c} &\multicolumn{1}{|c}{0.1462-0.0242i}&  \multicolumn{1}{|c}{-0.1236+0.3672i}&\multicolumn{1}{|c||}{0.0375+0.1035i}  \\
\hline
{Fig. 3d} &\multicolumn{1}{|c}{0.1106-0.1581i}&  \multicolumn{1}{|c}{-0.0591-0.6055i}&\multicolumn{1}{|c||}{0.2586-0.0054i}  \\
\hline
\end{tabular}
\label{S-opt-fig3}
\end{table}

The optical $S$ matrix was substituted in Eq. (\ref{w(g) beta1 N=1}), which is the theoretical 
prediction for the conductance probability distribution
$w_{\langle S \rangle}^{(1)}(T)$, giving, after normalizing to the total number of cases,
the results shown as the continuous lines in Fig. 3.
In other words, the parameters on which the theoretical results depend, i.e., 
the optical $S$ matrix elements, were not obtained by a variation procedure designed to find a 
best fit, but rather extracted from the numerical experiment.
In panels (a) and (b) the optical  $S$ matrix is very close to zero,
indicating a negligible presence of direct processes, so that
the continuous curve in both cases is practically given by
Eq. (\ref{w(g) beta1 N=1 S0}):
we mainly have long lived states in these structures.
The elements of the optical  $S$ matrix grow larger as we proceed to the remaining panels.

\begin{figure}[h]
\epsfig{file=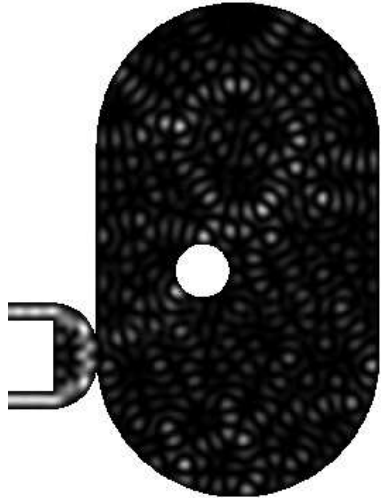,width=5cm,clip=}
\caption{{\footnotesize 
The square of the scattering wave function, i.e.,
$|\psi_E({\bf r})|^2$, for a fixed energy. 
The geometric structure consists of a small stadium coupled to a larger one.
The geometry is the same as that shown in Fig. \ref{w(T)22-23}(d).
We interpret the concentration of the wave function along the wall of the small cavity as a whispering gallery mode.
The larger stadium provides a ``sea" of fine-structure resonances.
}}
\label{wgm}
\end{figure}

The agreement between theory and numerical experiments is very good for (a) and (b).
One observes some deviations in panel (c); the deviations are largest for panel (d), where the optical, or direct, transmission, is largest.
In (c) the direct path between the two waveguides is obvious.
In (d) the direct path occurs inside
the smaller stadium, which supports whispering gallery modes, while the larger stadium provides a ``sea" of fine-structure states.
This effect is seen in Fig. \ref{wgm}, where a plot of the square of the scattering wave function
for a fixed energy $E$, i.e.,
$|\psi_E({\bf r})|^2$,
exhibits a concentration along the wall of the smaller stadium.
Indeed, the reason for attaching a small stadium to a large one in 
Fig. \ref{w(T)22-23}(d) is 
precisely that the whispering gallery modes which have been seen in calculations for small cavities, as in Refs. [\onlinecite{ingrid(2002),wgm2}], 
are gradually destroyed if the size of the cavity is increased, because of the long way the whispering gallery mode would have to traverse
(for more details the reader is referred to Ref. \onlinecite{ingrid_II}). This 
seems 
to be the reason why no effects from WGM are seen in the stadium in Fig. 
\ref{w(T)22-23}(b).

\begin{figure}[h]
\epsfig{file=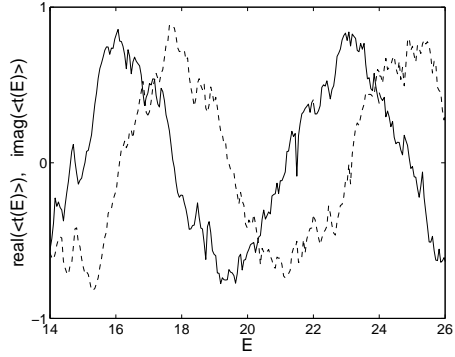,width=6cm,clip=}
\caption{{\footnotesize
The real (solid line) and imaginary (dashed line) parts of the 
ensemble expectation value 
$\langle t(E) \rangle$ (an element of the ``local" optical matrix)
as a function of energy, for the structure shown in
Fig. \ref{w(T)22-23} (d).
The question is whether the variation of these quantities
inside the energy interval $\Delta E = \left[ 22,23 \right]$
is responsible for the discrepancy seen in Fig. \ref{w(T)22-23} (d).
}}
\label{<t>(d)}
\end{figure}

\begin{figure}[h]
\epsfig{file=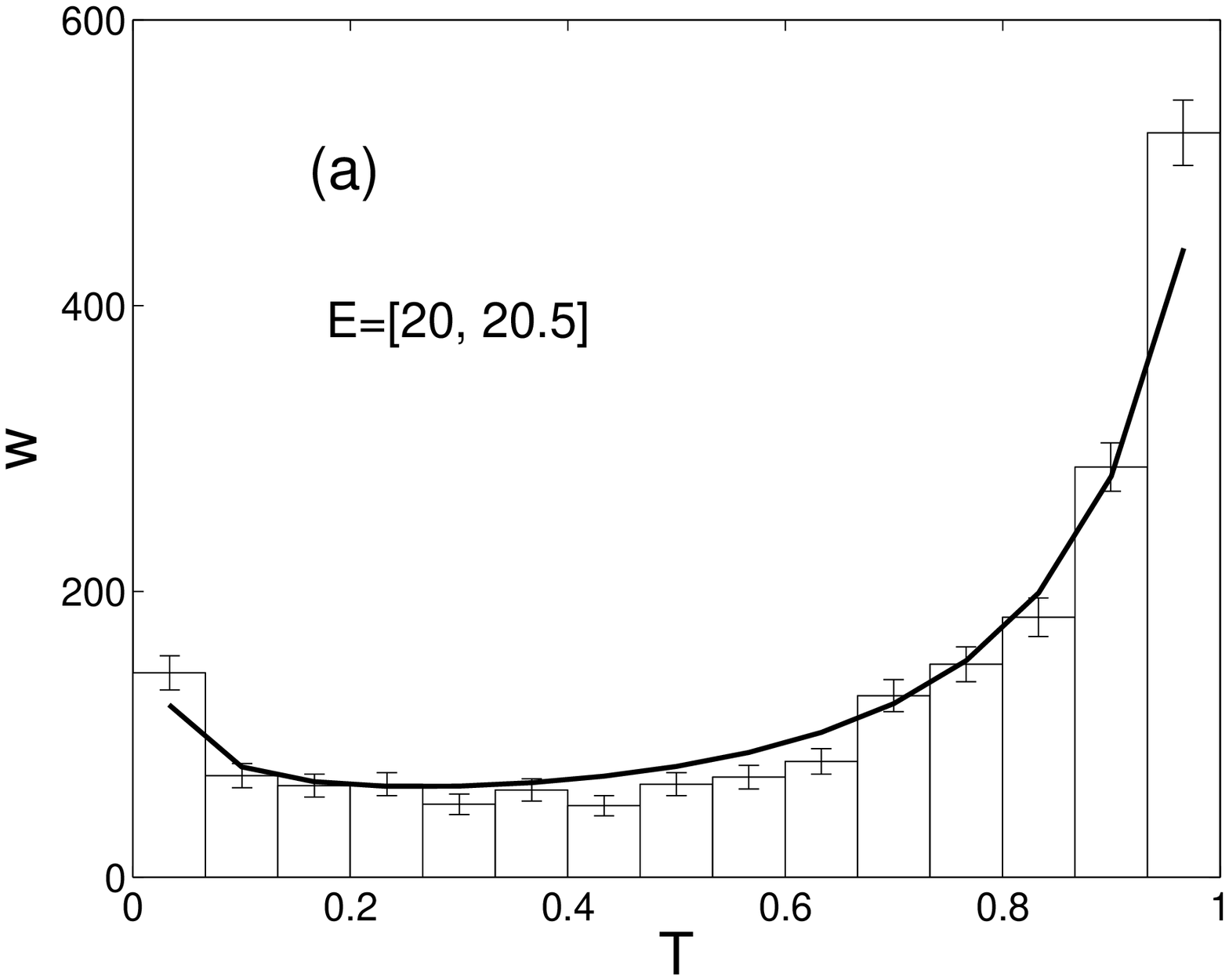,width=6cm,clip=}
\epsfig{file=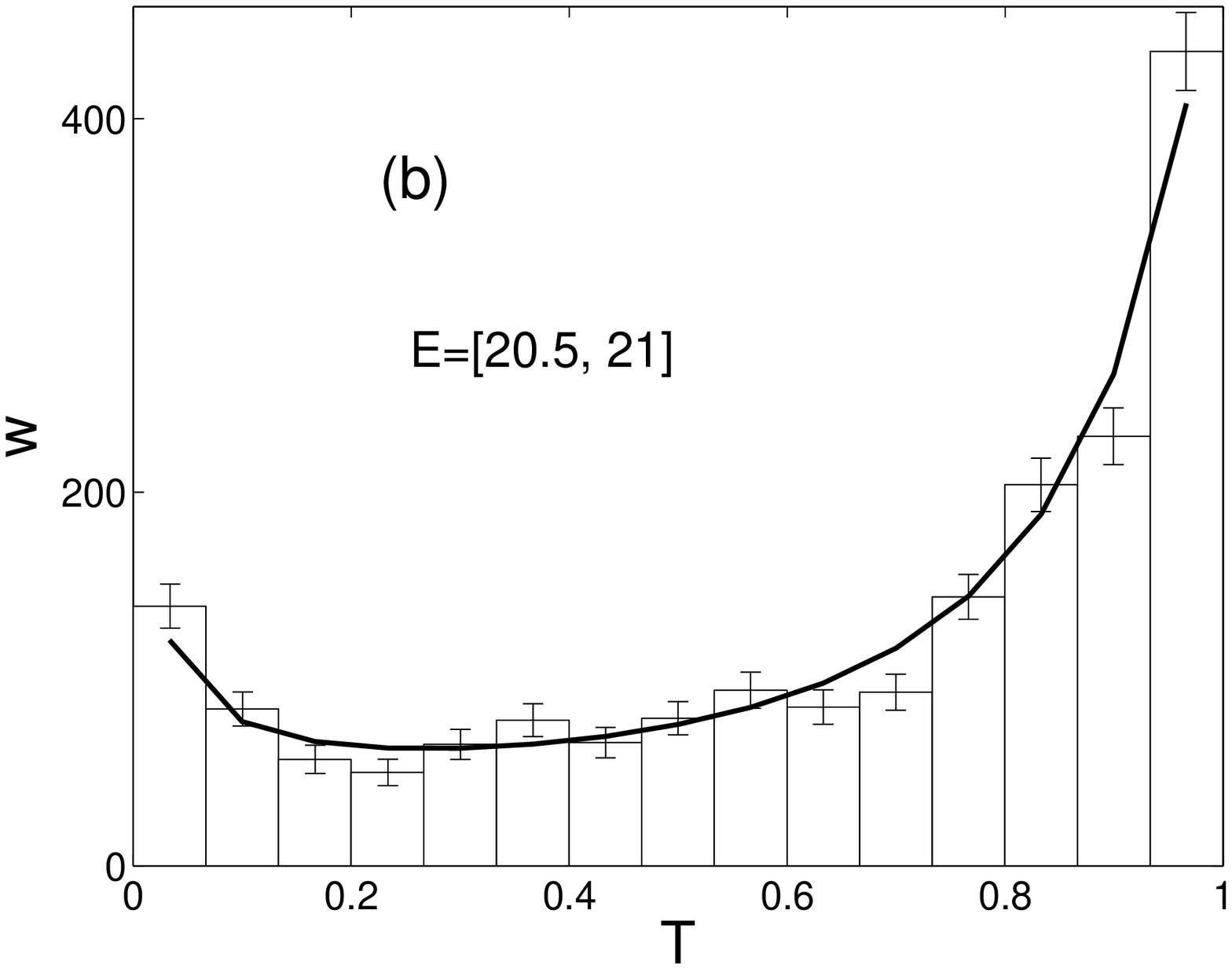,width=6cm,clip=}
\epsfig{file=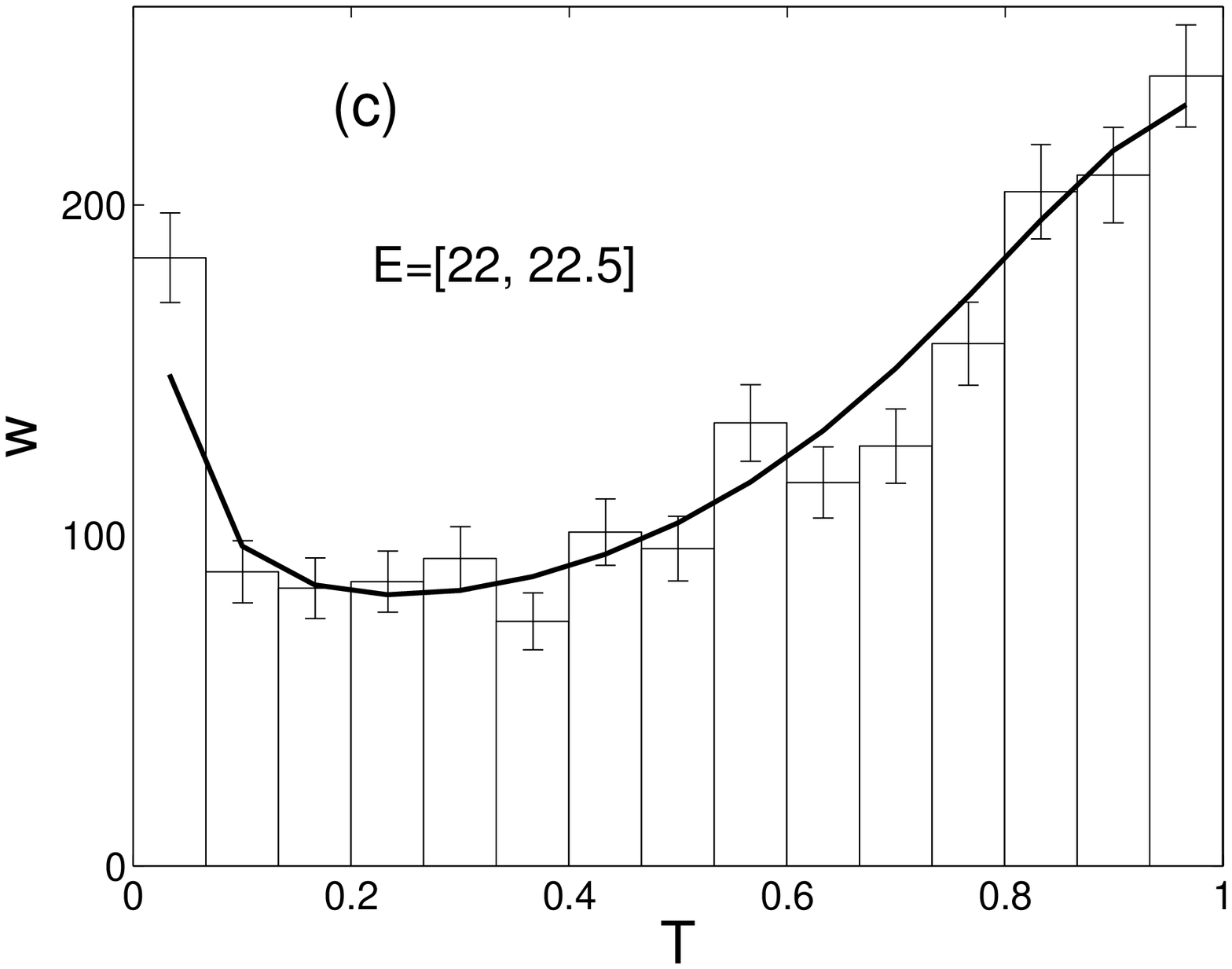,width=6cm,clip=}
\epsfig{file=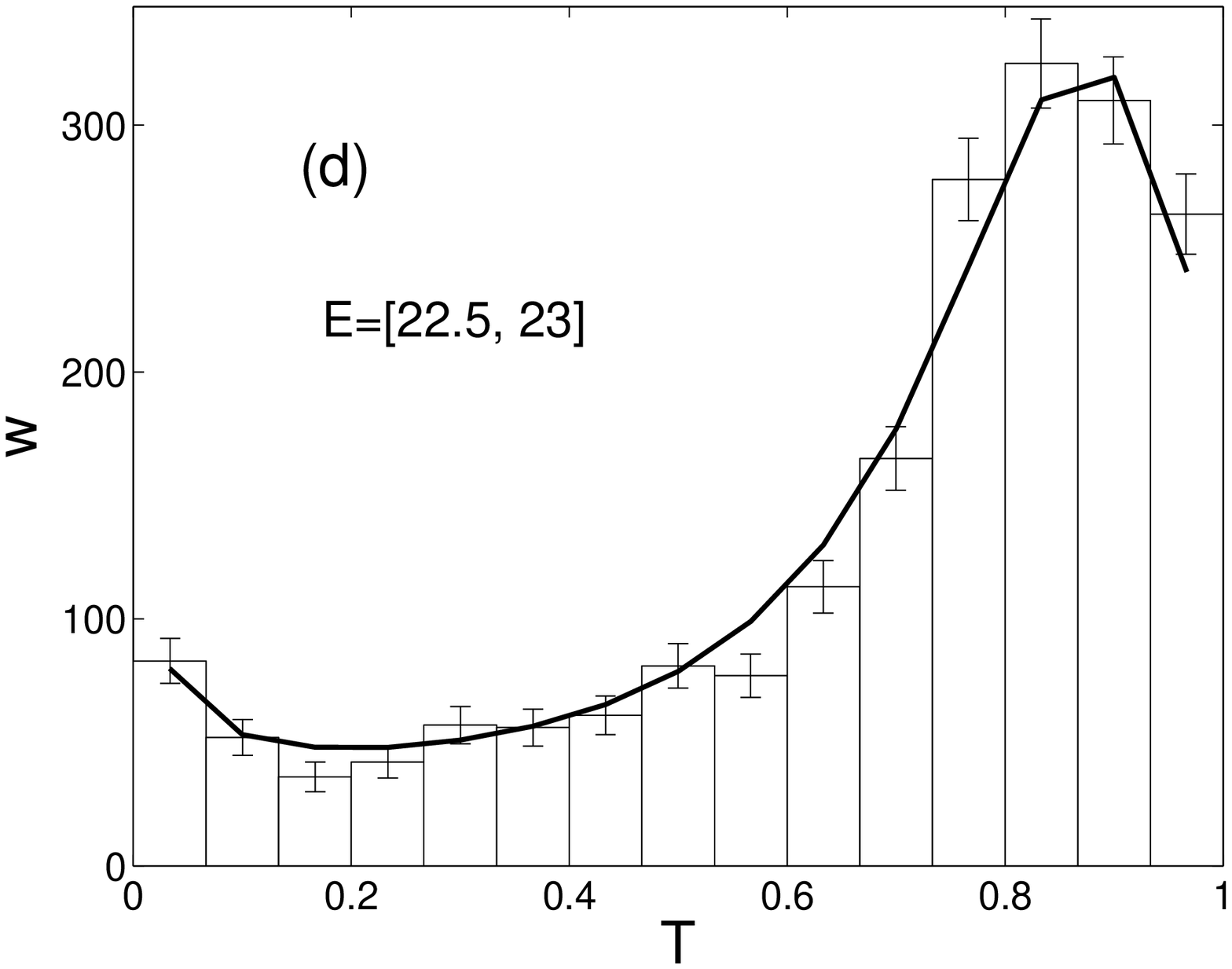,width=6cm,clip=}
\caption{{\footnotesize
The distribution $W$ of the conductance,
normalized to the total number of cases,
for the same structure as in Fig. \ref{w(T)22-23} (d), but using energy intervals, shown in each panel, twice as small for the construction of the histograms.
Panels (c) and (d) show the same data of Fig. \ref{w(T)22-23} (d),
but analyzed inside each of the two subintervals.
Panels (a) and (b) show the data for two other similar subintervals.
The agreement with theory is very good.
}}
\label{w(T)20 20.5}
\end{figure}

We wish to investigate the case of Fig. \ref{w(T)22-23} (d) further. 
For the optical $S$ matrix extracted from the data, the probability of small
$T$'s predicted by the theory is larger than that found in the numerical
simulation, and vice versa for $T \sim 1$.
This effect is not a statistical fluctuation, but rather a systematic discrepancy,
as it was observed in most cases where the transmission part of the optical $S$ matrix
is as large as that occurring in Fig. \ref{w(T)22-23}(d).
To find the origin of the discrepancy we have to realize that,
in order to apply our model meaningfully, an energy interval $\Delta E$ over which we do statistics
must be such that the ``local" optical matrix $\langle S(E) \rangle$ is ``reasonably constant" 
across it, while, at the same time, such an interval should contain many fine structure resonances:
in other words, the notion of ``stationarity" should be approximately valid.
Figure \ref{<t>(d)} shows the ensemble expectation value
$\langle t(E) \rangle$
of the transmission amplitude $t(E)$ as a function of the energy $E$, for the structure shown in Fig. \ref{w(T)22-23} (d). 
Although this quantity is certainly not expected to be literally constant, 
the question is whether its variation across the energy interval $\Delta E = \left[ 22,23 \right]$
(used in  Fig. \ref{w(T)22-23} (d))
can be considered to be the cause of the discrepancy that we have seen between theory and numerical experiment:
the following results support our belief that the answer to this question is positive.
Figure \ref{w(T)20 20.5} shows the conductance distribution for the same structure of
Fig. \ref{w(T)22-23} (d), obtained using a number of energy intervals 
twice as small. For instance, panels (c) and (d) correspond to the two
subintervals in which the original one, $\Delta E = \left[ 22,23 \right]$, was divided.
Panels (a) and (b) show the data for two other similar subintervals.
We observe that the agreement is now very good.

\subsubsection{Statistical properties of the shot-noise power spectrum}
\label{num shot N=1}

The shot-noise power
spectrum at zero temperature of Eq. (\ref{<P> 2}) was calculated, 
both numerically as well as from our theoretical model,
for the same structures shown in Figs. \ref{w(T)22-23} and \ref{w(T)20 20.5}.
For one channel, $N=1$, Eq. (\ref{<P> 2}) simplifies to
\begin{equation}
\frac{\langle P \rangle}{\langle P_P \rangle}
=\frac{\langle T(1-T) \rangle}{\langle T \rangle} ,
\label{shot N1}
\end{equation}
so that in this case we do not have more information than that contained in the conductance distribution.
However, for completeness, we present the results in 
Table \ref{shot noise N=1}.

Notice that the results in the first two rows of the table, i.e., those arising  from Figs. \ref{w(T)22-23}(a), (b)
(with an optical $S$ close to zero), compare
well with the prediction of Eq. (\ref{<P> beta1}) for $N=1$, i.e.,
$\langle P \rangle / \langle P_P \rangle = 0.4$,
and with Fig. \ref{fanobeta2N1} for $\beta=1$ and $\langle S \rangle =0$.

Notice also that from row 1(or 2) to row 4 of the table the optical transmission increases and
$\langle P \rangle / \langle P_P \rangle$ decreases, in qualitative agreement with the result of
Fig. \ref{fanobeta2N1} for $\beta=1$,  $\langle r \rangle$ literally equal to zero and increasing $\langle t \rangle$.

\begin{table}[h]
\caption{\footnotesize The shot-noise power spectrum of Eq. (\ref{shot N1}), 
$N=1$}
\begin{tabular}{||ccc||}
\hline\hline
{Case} &  \multicolumn{1}{|c}{Numerical} &\multicolumn{1}{|c||} {Theoretical}  \\
\hline\hline
{Fig. 3a} &\multicolumn{1}{|c}{$0.3961 \pm 0.0071$}&\multicolumn{1}{|c||}{0.4000}  \\
\hline
{Fig. 3b} &\multicolumn{1}{|c}{$0.3813 \pm 0.0084$}&\multicolumn{1}{|c||}{0.4000}  \\
\hline
{Fig. 3c} &\multicolumn{1}{|c}{$0.2979 \pm 0.0013$}&\multicolumn{1}{|c||}{0.3251}  \\
\hline
{Fig. 3d} &\multicolumn{1}{|c}{$0.2315 \pm 0.0026$}&\multicolumn{1}{|c||}{0.2438}  \\
\hline
{Fig. 6a} &\multicolumn{1}{|c}{$0.1765 \pm 0.0031$}&\multicolumn{1}{|c||}{0.1959}  \\
\hline
{Fig. 6b} &\multicolumn{1}{|c}{$0.1972 \pm 0.0041$}&\multicolumn{1}{|c||}{0.2001}  \\ 
\hline
{Fig. 6c} &\multicolumn{1}{|c}{$0.2576 \pm 0.0042$}&\multicolumn{1}{|c||}{0.2587}  \\ 
\hline
{Fig. 6d} &\multicolumn{1}{|c}{$0.2104 \pm 0.0029$}&\multicolumn{1}{|c||}{0.2187}  \\ 
\hline
\end{tabular}
\label{shot noise N=1}
\end{table}

\subsection{The two-channel case, N=2}
\label{num N2}

In units such that $\hbar^2/2mW^2 =1$, two open channels ($N=2$) occur in the energy interval
$E \; \epsilon \; [4\pi^2, 9\pi^2]\approx [40, 90]$.
In view of the experience gained in the one-channel case described above,
the energy interval $\Delta E$ was literally reduced to a point, and
the statistical properties of the conductance
and the shot-noise power spectrum were studied across the ensemble for a fixed energy $E$:
more specifically, 200 samples were collected at $E=75$, varying, just as before, 
the position of the obstacle inside the cavity.

\subsubsection{Statistical properties of the conductance}

\begin{widetext}
\begin{figure}[h]
\epsfig{file=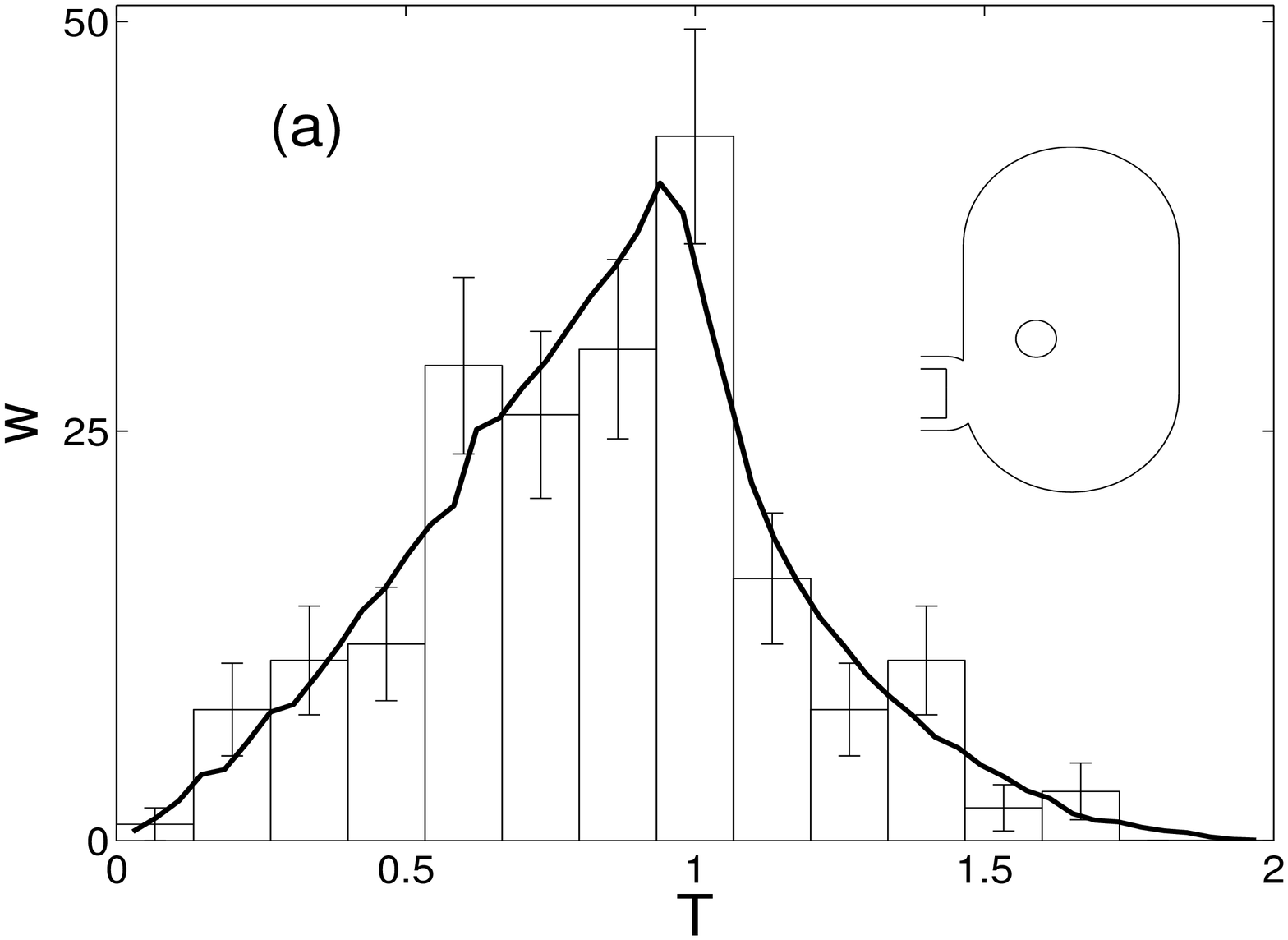,width=6cm,clip=}
\epsfig{file=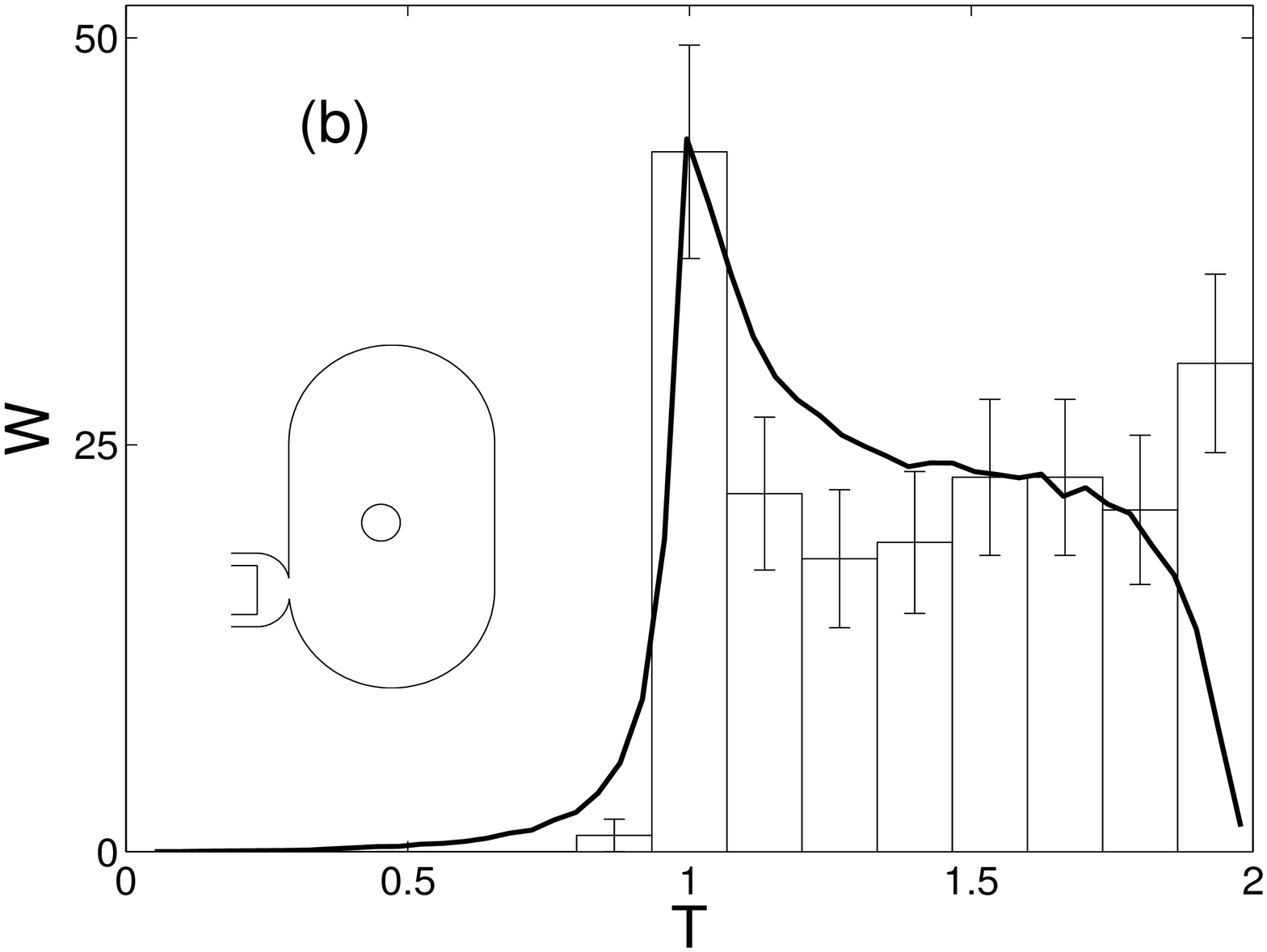,width=6cm,clip=}
\caption{{\footnotesize
The distribution $W$ of the conductance,
normalized to the total number of cases,
for the structures shown in the insets and for two open channels ($N=2$):
the structures in panels (a) and (b)
correspond to those shown in panels (a) and (d) of Fig. \ref{w(T)22-23}, respectively.
The structure in (a) consists of a Bunimovich stadium connected to two waveguides directly,  
while in (b) the connection is done through a smaller half stadium.
The histograms were obtained from a numerical solution of the Schr\"odinger equation for these structures at the energy $E=75$, and constructing an ensemble of 200 positions of the obstacle. 
The optical $S$ matrix was extracted from the data and used to obtain, from Poisson's kernel, the theoretical distributions;
these were computed at 50 points in the interval $0<T<2$, which were then joined to obtain the continuous curves.
The trends shown by the numerical distributions are well bescribed by theory.
}}
\label{w(T) 75}
\end{figure}
\end{widetext}

The numerical simulation was done by solving numerically the Schr\"odinger equation
for the structures shown in Figs. \ref{w(T)22-23}(a) and \ref{w(T)22-23}(d).
The theoretical prediction for the conductance distribution $w(T)$ 
was obtained using Eq. (\ref{wT N=2}), which in turn makes use of
Eq. (\ref{w(tau) beta1 N=2}) and the equations given in the appendix: 
the integrations occurring in the equations of the appendix were performed numerically
using a Monte Carlo method (Metropolis algorithm).
The optical $S$ matrix $\langle S(E) \rangle$ that was used was extracted from the data at $E=75$ and across the ensemble; 
it is given in the following equation for Fig. \ref{w(T) 75}(a)

\begin{eqnarray}
\langle S(E)  \rangle
=\left[
\begin{tabular}{c|c}
$\begin{array}{cc}
-0.0312 + 0.0259i  &  0.0805 - 0.0393i   \\
 0.0805 - 0.0393i  & -0.1600 + 0.2470 i
\end{array}
$
& 
$\begin{array}{cc}
-0.1391 + 0.0599 i  &  0.1186 + 0.0908 i \\
-0.0763 - 0.0091 i  &  0.0982 + 0.0056 i
\end{array}
$
\\ \hline
$\begin{array}{cc}
-0.1391 + 0.0599 i  &  -0.0763 - 0.0091 i    \\
 0.1186 + 0.0908 i  &   0.0982 + 0.0056 i
\end{array}
$
& 
$\begin{array}{cc}
0.1032 + 0.0091 i   &    0.0357 - 0.0556 i \\
0.0357 - 0.0556 i   &    0.0723 + 0.0764 i
\end{array}
$
\end{tabular}
\right] ,
\nonumber \\
\label{<S>7a}
\end{eqnarray}
while for the cavity shown in Fig. \ref{w(T) 75}(b), the optical $S$ matrix
is given by
\begin{eqnarray}
\langle S(E)  \rangle
=\left[
\begin{tabular}{c|c}
$\begin{array}{cc}
0.1536 - 0.1256 i  &  0.0469 + 0.0313 i   \\
0.0469 + 0.0313 i  &  0.0777 - 0.0255 i
\end{array}
$
& 
$\begin{array}{cc}
 0.0703 - 0.4275 i  &  0.0743 - 0.2620 i \\
 0.0204 + 0.2628 i  &  0.7589 + 0.1623 i
\end{array}
$
\\ \hline
$\begin{array}{cc}
0.0703 - 0.4275 i  & 0.0204 + 0.2628 i  \\
0.0743 - 0.2620 i  & 0.7589 + 0.1623 i
\end{array}
$
& 
$\begin{array}{cc}
-0.0135 - 0.1261 i &  0.0929 - 0.0125 i \\
 0.0929 - 0.0125 i & -0.0452 - 0.0195 i
\end{array}
$
\end{tabular}
\right] .
\nonumber \\
\label{<S>7b}
\end{eqnarray}
The blocks in the two previous equations indicate the optical transmission and reflection matrices, as in Eq. (\ref{S opt N2}).
Notice that $\langle S(E) \rangle \approx 0$ in Eq. (\ref{<S>7a}), while Eq. (\ref{<S>7b}) shows a large direct transmission.

The results given in Fig. \ref{w(T) 75} show a strong effect of
direct processes on the conductance distribution, 
whose trends are well described by Poisson's kernel.

\subsubsection{Statistical properties of the shot-noise power spectrum}

The theoretical predictions 
for the average of the shot-noise power spectrum of Eq. (\ref{P/PP N2})
were compared with the results of the
numerical simulation.
Notice that in the two-channel case the statistics of the shot-noise power spectrum
gives information which is not contained in the conductance distribution.
The comparison is shown in Table \ref{shot noise N=2} for the same cases denoted as (a) and (b)
in Fig. \ref{w(T) 75}. We note that case (a), whose $\langle S \rangle$ is
close to zero, compares reasonably
well with the theoretical result 
[$\langle P \rangle / \langle P_P \rangle = 0.31$]
of Eq. (\ref{<P> beta1}) for $N=2$ and $\langle S \rangle =0$.

\begin{table}[h]
\caption{\footnotesize The shot-noise power spectrum of Eq. (\ref{P/PP N2}),
  $N=2$ }
\begin{tabular}{||ccc||}
\hline\hline
{Case} &  \multicolumn{1}{|c}{Simulation} &\multicolumn{1}{|c||} {Theoretical}  \\
\hline\hline
{(a)} &\multicolumn{1}{|c}{$0.2959 \pm 0.0030$}&\multicolumn{1}{|c||}{0.3300}  \\
\hline
{(b)} &\multicolumn{1}{|c}{$0.1144 \pm 0.0011$}&\multicolumn{1}{|c||}{0.1200}  \\
\hline
\end{tabular}
\label{shot noise N=2}
\end{table}

\section{Conclusions and Discussion}
\label{discussion}

The statistical properties of the dc conductance in chaotic cavities have been investigated in the past in the framework of the maximum-entropy model described in the previous sections. 
Within the same framework, in the present paper we have gone further by studying, in addition to the conductance, the zero-frequency shot-noise power spectrum.
The shot noise is a more complicated quantity than the conductance, in the sense that 
it involves electron correlations due to the Pauli principle.
We have been particularly interested in the effect that direct processes consisting of whispering gallery modes have on the conductance and on the shot-noise power;
these modes were promoted by choosing properly the structure of the cavities and
the position of the leads. 
This kind of direct processes were, in fact, avoided in previous publications by some of the present authors. 
For the two symmetries ($\beta =1,2$) studied here we have found that the
ratio $\langle P \rangle^{(\beta)}/\langle P_P\rangle^{(\beta)}$, as a function
of the number of channels for $\langle S \rangle =0$, is 
larger 
for $\beta=1$ than for $\beta=2$, indicating that small 
values of the transmission eigenvalues are favored by time-reversal symmetry.

We have found that the agreement between the theoretical predictions and the
results of computer simulations performed for one and two open channels
is generally good.
However, the systematic discrepancies that we have observed lead us to revise the notions under which our model has been constructed.

Indeed, the maximum-entropy model described in Sec. \ref{PK} relies on a number of assumptions.
For instance, the extreme idealization is made of regarding $S(E)$ as a ``stationary random (matrix) function"
of energy.
As a consequence, the optical matrix $\langle S(E) \rangle$ is constant with energy and 
the characteristic time associated with direct processes is literally zero.
The property of {\em stationarity} allows defining the notion of {\em ergodicity} which, 
together with {\em analitycity}, gives
the {\em reproducing property}, Eq.  (\ref{reprod}),
which is essential for the definition of Poisson's kernel (PK)
of Eq. (\ref{poisson}).

Needless to say, in realistic dynamical problems stationarity is only approximately fulfilled, so that
one has to work with energy intervals $\Delta E$
across which the ``local" optical matrix $\langle S(E) \rangle$ is 
{\em approximately constant},
while, at the same time, such intervals should contain many fine-structure resonances.
This compromise can actually be realized in Nuclear Physics,
where the optical $\langle S \rangle$ arises from the tail of many
distant resonances or
from a single-particle resonance that lies so far away in the complex-$E$ plane
to act as a smooth background on top of the fine-structure compound-nucleus
resonances: hence the huge separation between the two time scales.
In contrast, as we saw in Sec. \ref{num}, such a compromise is difficult to fulfill for the physical systems studied here:
this we believe to be the origin of the discrepancies observed between theory and numerical simulations.
(Indeed, discrepancies similar to the ones that we have observed in this paper 
were already there, to a certain extent, in Refs. 
[\onlinecite{baranger-mello(epl),
baranger-mello(wrm)}], but were overlooked at that time.)
In the present paper we give evidence that reducing $\Delta E$ literally to a point
and collecting data over an ensemble constructed by changing the position of the obstacle inside the cavity,
the agreement between theory and experiment is significantly improved,
being excellent in several cases.
In other words, {\em PK gives a good description of the statistics of the data taken across the ensemble}.

It is interesting to remark that also in Ref. 
[\onlinecite{baranger-mello(epl)}] cases had been found
in which stationarity obviously did not hold.
Energy averages were out of the question in those cases, so that an ensemble was generated by adding ``noise"
along the wall: it was found that PK gave an excellent description of the data collected
across the ensemble at a fixed $E$.
This point was merely indicated at that time and no results were published.

Thus the results shown in the present paper give evidence that PK is valid beyond the situation where it was originally derived,
which required the properties of analyticity, stationarity and ergodicity, plus a maximum-entropy ansatz.
It is as though the reproducing property of Eq. (\ref{reprod}) were valid even in the absence of
stationarity and ergodicity (analyticity is always there, of course).
Even at the present moment we are unable to give an explanation of this fact.
A few remarks are in order in connection with this point.

Let us take the invariant measure $d\mu (S_0)$
of Sec. \ref{PK}
as a model for the description of scattering by a chaotic cavity
described by the scattering matrix $S_0$
and assumed to have ideal coupling to the leads.
Brouwer has shown (see Ref. [\onlinecite{piet-PK}], Sec. V) that when such a chaotic cavity
is coupled to the leads by means of a tunnel barrier (non-ideal coupling)
described by a fixed scattering matrix $S_1$, say,
the resulting $S$, constructed using the combination law of $S_0$ and $S_1$,
is distributed according to PK.
Brouwer's proof, being essentially a change of variables from $S_0$ to the final $S$, 
does not require stationarity, or ergodicity, or the maximum-entropy ansatz;
however, it neglects evanescent modes between the barrier and the cavity. 
In other words, the reproducing property, which is fulfilled identically
for the invariant measure \cite{mello-pereyra-seligman,mello-seligman},
is not destroyed by the presence of the tunnel barriers.
The latter certainly give rise to a nonzero $\langle S \rangle$, so that
the direct processes described by this $\langle S \rangle$,
being produced by the tunnel barriers, take place outside the cavity
(see Fig. 2 in Ref. \onlinecite{piet-PK}).
In contrast, when direct processes take place inside the system, it is not
possible, in general, to write the total $S$ as the combination of an $S_0$
and a {\it fixed} $S_1$, as required by Brouwer's analysis.
Take, for instance, the system shown in Fig. \ref{w(T)22-23}(d).
If we had, say, a long ``neck" between the small cavity and the big one, then we could define scattering matrices $S_1$ for the former and $S_0$ for the latter and
combine them, disregarding evanescent modes, to obtain the total 
scattering matrix $S$.
However, this is not the case for the actual system under study.
As an approximation, we might think of assigning to the small and big cavities of the system of Fig. \ref{w(T)22-23}(d) the scattering matrices $S_1$ and $S_0$, respectively, that would occur if we added the neck between the two;
the total $S$ obtained by combining these open-channel $S_1$ and $S_0$ would represent an approximation to the actual problem; 
however, we are not in a position to know how close this approximation 
would be to the exact solution:
we leave this open question for future investigation.
Once again we seem to find that the valididty of PK for the systems studied in the previous section goes beyond the domain in which Brouwer's result was derived.

Brouwer has also shown \cite{piet-PK} that PK for the $S$ matrix can be obtained from a
Lorentzian ensemble of Hamiltonians with an arbitrary number of levels $M$. 
In the limit $M \to \infty$ the Lorentzian ensemble becomes equivalent to a Gaussian ensemble. 
In this limit, in which we believe that the Gaussian ensemble
describes a chaotic cavity, the problem becomes once again stationary in energy.

It thus seems that a derivation of PK 
--or at least of the reproducing property-- 
for chaotic cavities with a general type of direct processes and in the absence of stationarity is, to our knowledge, still missing.

When this work was completed, the present authors became aware of a study of
the shot noise problem by D. Savin et al., \cite{savin} and P. Braun et
al. \cite{braun} in which results similar to those of our 
Sec. \ref{N arb} have been obtained.

\acknowledgments

One of the authors (P.A.M.) whishes to acknowledge the hospitality of the
Max-Planck-Institut f\"ur Physik komplexer Systeme (MPI-PKS) in 
Dresden, for making possible a long-term visit during which the present work 
could be almost completed. He also acknowledges financial support by Conacyt,
M\'exico, under Contract No. 42655.   
He also wishes to thank H. U. Baranger, C. Lewenkopf, M. Mart\'inez and  
T. H. Seligman for useful discussions.
E.N.B. and V.A.G. are also grateful to the MPI-PKS for its hospitality during their
stay in Dresden.

\appendix

\section{The polar representation, the invariant measure and some 
statistical distributions for one and two channels}
\label{app}

For completeness, we present the explicit parametrization of the $S$ matrix in the polar representation for $N=1$ and $N=2$ and some of its applications.

\subsection{The one-channel case, $N=1$.}

We write the two-dimensional $S$ matrix in the polar representation as
\begin{equation}
S =
\left[
\begin{array}{cc}
-\sqrt{1-\tau}\;e^{i(\alpha + \gamma)} &  \sqrt{\tau}\;e^{i(\alpha + \delta)}\\
 \sqrt{\tau}\;e^{i(\beta + \gamma)} & \sqrt{1-\tau}\;e^{i(\beta + \delta)}
\end{array}
\right]
=\left[
\begin{array}{cc}
r &  t'   \\
t & r'
\end{array}
\right]
\label{S polar N=1}
\end{equation}
and the optical $\langle S \rangle$ as
\begin{equation}
\langle S \rangle =
\left[
\begin{array}{cc}
\langle r \rangle  & \langle t' \rangle  \\
\langle t \rangle  & \langle r' \rangle
\end{array}
\right] ,
\label{S opt}
\end{equation}
where the various entries are complex numbers.
For $\beta =1$ one has the restrictions $\gamma = \alpha$ and $\delta = \beta $.
The distribution of the conductance $T=\tau$ for $\beta =1$ can be reduced to quadratures, with the result given in the text, Eq. (\ref{w(g) beta12 N=1}).

\subsubsection{The two-channel case, $N=2$.}

The expressions given below are used in the present work when carrying out the numerical computations; since these are performed for the orthogonal case, $\beta =1$, we restrict ourselves to this universality class.
For $\beta =1$ we write the four-dimensional $S$ matrix in the polar representation as
\begin{equation}
S =
\left[
\begin{array}{cc}
-v^{(1)}\sqrt{1-\tau}\;[v^{(1)}]^T & v^{(1)} \sqrt{\tau}\;[v^{(2)}]^T\\
 v^{(2)}\sqrt{\tau}\;[v^{(1)}]^T & v^{(2)}\sqrt{1-\tau}\;[v^{(2)}]^T
\end{array}
\right]
=\left[
\begin{array}{cc}
r &  t'   \\
t & r'
\end{array}
\right] \; .
\label{S 2}
\end{equation}
The reflection and transmission matrices $r$, $t$, etc., are two dimensional.
The matrix $\tau $ is two dimensional and diagonal: $\tau _{ab}= \tau _a \delta _{ab}$,
with $0\leq \tau _a \leq 1 $.
The matrices $v^{(1)}$ and $v^{(2)}$ are two-dimensional unitary matrices which can be written as
\begin{equation}
v^{(i)} =
\left[
\begin{array}{cc}
-\sqrt{1-\tilde{\tau}^{(i)}}\;e^{i(\alpha^{(i)} + \gamma^{(i)})} &  \sqrt{\tilde{\tau}^{(i)}}\;e^{i(\alpha^{(i)} + \delta^{(i)})}\\
\sqrt{\tilde{\tau}^{(i)}}\;e^{i(\beta^{(i)}+\gamma^{(i)})}
& 
\sqrt{1-\tilde{\tau}^{(i)}}\;e^{i(\beta^{(i)} + \delta^{(i)})}
\end{array}
\right ] .
\label{vi 2}
\end{equation}
The optical $\langle S \rangle$ is written as
\begin{equation}
\langle S \rangle =
\left[
\begin{array}{cc}
\langle r\rangle  & \langle t' \rangle \\
\langle t \rangle & \langle r' \rangle
\end{array}
\right] ,
\label{S opt N2}
\end{equation}
where the various entries are two-dimensional matrices.

The invariant measure for $v^{(i)}$ is given by
\begin{equation}
d\mu (v^{(i)})
= d\tilde{\tau}^{(i)} \; \frac{d\alpha ^{(i)} d\beta ^{(i)} d\gamma ^{(i)}  d\delta ^{(i)}}{(2\pi )^4},
\label{dmu vi}
\end{equation}
with the range of variation of the parameters
\begin{subequations}
\begin{eqnarray}
0&\leq& \alpha ^{(i)} , \; \beta ^{(i)} , \; \gamma ^{(i)} , 
\; \delta ^{(i)} \leq 2\pi ,
\\
0&\leq& \tilde{\tau}^{(i)} \leq 1
\end{eqnarray}
\label{range}
\end{subequations}
and is normalized as $\int d\mu (v^{(i)}) = 1$.

The joint probability distribution of 
$\tau _1, \tau _2$ is given
in Eq. (\ref{w(tau) beta1 N=2}) of the text: it is a 10-dimensional integral,
with $d\mu (v^{(i)})$ given in Eq. (\ref{dmu vi}),
the range of variation of the parameters being specified by (\ref{range}).

\end{document}